\input harvmac
\input amssym.tex

\def\newdate{Bures-sur-Yvette, 8/5/2004}


\def\a{\alpha}
\def\b{\beta}  
\def\g{\gamma}
\def\l{\lambda}
\def\d{\delta}
\def\e{\epsilon}
\def\t{\theta}

\def\L{\Lambda}

\def\p{\partial}
\def\pa{\partial}
\def\demi{{1\over 2}}


  \lref\OoguriPS{ 
  H.~Ooguri, J.~Rahmfeld, H.~Robins and J.~Tannenhauser, 
  JHEP {\bf 0007}, 045 (2000) 
  [hep-th/0007104]. 
  } 

  \lref\SYM{ 
  W.~Siegel, 
  Phys.\ Lett.\ B {\bf 80}, 220 (1979)\semi 
  S.~J.~Gates and S.~Vashakidze, 
  Nucl.\ Phys.\ B {\bf 291}, 172 (1987)\semi 
  B.~E.~Nilsson, 
  GOTEBORG-81-6\semi 
  B.~E.~Nilsson, 
  Class.\ Quant.\ Grav.\  {\bf 3}, L41 (1986);  
   M.~Cederwall, B.~E.~Nilsson and D.~Tsimpis, 
  JHEP {\bf 0106}, 034 (2001) 
  [hep-th/0102009];  
  M.~Cederwall, B.~E.~Nilsson and D.~Tsimpis, 
  JHEP {\bf 0107}, 042 (2001) 
  [hep-th/0104236]. 
  } 

  \lref\har{ 
  J.~P.~Harnad and S.~Shnider, 
  Commun.\ Math.\ Phys.\  {\bf 106}, 183 (1986). 
  } 

  \lref\purespinors{\'E. Cartan, {\it Lecons sur la th\'eorie des spineurs},  
  Hermann, Paris (1937)\semi 
  C. Chevalley, {\it The algebraic theory of Spinors},  
  Columbia Univ. Press., New York\semi 
   R. Penrose and W. Rindler,  
  {\it Spinors and Space-Time}, Cambridge Univ. Press, Cambridge (1984)  
  \semi 
  P. Budinich and A. Trautman, {\it The spinorial chessboard}, Springer,  
  New York (1989). 
  } 

  \lref\GS{M.B. Green, J.H. Schwarz, and E. Witten, {\it Superstring Theory,}  
   vol. 1, chapter 5 (Cambridge U. Press, 1987).  Our  
  $x^m$ differs  by a sign, and our $\theta^\a=\sqrt{2} \, \theta^\a_{\rm GSW}$. 
  } 

  \lref\csm{W. Siegel, 
  Nucl. Phys. B263 (1986) 93;   
  W.~Siegel,   Phys. Rev. D 50 (1994), 2799. 
  }    

   \lref\fms{D. Friedan, E. Martinec and S. Shenker,  
  Nucl. Phys. B271 (1986) 93.} 

  \lref\BerkovitsFE{ 
  N.~Berkovits, 
  JHEP { 0004}, 018 (2000) 
  [hep-th/0001035]. 
  } 

  \lref\BerkovitsPH{ 
  N.~Berkovits and B.~C.~Vallilo, 
  JHEP { 0007}, 015 (2000) 
  [hep-th/0004171]. 
  } 

  \lref\BerkovitsNN{ 
  N.~Berkovits, 
  JHEP { 0009}, 046 (2000) 
  [hep-th/0006003]. 
  } 

  \lref\BerkovitsWM{ 
  N.~Berkovits, 
  Int.\ J.\ Mod.\ Phys.\ A { 16}, 801 (2001) 
  [hep-th/0008145]. 
  } 

  \lref\BerkovitsYR{ 
  N.~Berkovits and O.~Chandia, 
  Nucl.\ Phys.\ B {\bf 596}, 185 (2001) 
  [hep-th/0009168]. 
  } 

  \lref\BerkovitsZY{ 
  N.~Berkovits, 
  Nucl.\ Phys.\ B {\bf 420}, 332 (1994) 
  [hep-th/9308129]. 
  } 

  \lref\BerkovitsUS{ 
  N.~Berkovits, 
  hep-th/0104247. 
  } 

  \lref\BerkovitsMX{ 
  N.~Berkovits and O.~Chandia, 
  hep-th/0105149. 
  } 

\lref\BerkovitsUE{
N.~Berkovits and P.~S.~Howe,
Nucl.\ Phys.\ B {\bf 635}, 75 (2002)
[hep-th/0112160].
}

\lref\berko{
N.~Berkovits,  
JHEP {\bf 0004}, 018 (2000); 
N.~Berkovits,  
Int.\ J.\ Mod.\ Phys.\ A {\bf 16}, 801 (2001)  
[hep-th/0008145]; 
N.~Berkovits, 
  JHEP { 0009}, 046 (2000) 
  [hep-th/0006003]; 
N.~Berkovits,
{\it ICTP lectures on covariant quantization of the superstring,}
[hep-th/0209059].
}

  \lref\howe{P.S. Howe, 
  Phys. Lett. B258 (1991) 141, Addendum-ibid.B259 (1991) 511\semi  
  P.S. Howe, 
  Phys. Lett. B273 (1991)  
  90.} 

\lref\GreenNN{
M.~B.~Green,
Phys.\ Lett.\ B {\bf 223}, 157 (1989).
}

\lref\tonin{ 
I. Oda and M. Tonin, 
Phys. Lett. B520 (2001) 398 [hep-th/0109051]\semi 
M.~Matone, L.~Mazzucato, I.~Oda, D.~Sorokin and M.~Tonin,
Nucl.\ Phys.\ B {\bf 639}, 182 (2002)
[hep-th/0206104].
}  

\lref\zwie{
B.~Zwiebach,
Nucl.\ Phys.\ B {\bf 390}, 33 (1993)
[hep-th/9206084].
}

\lref\grassi{
P.~A.~Grassi, G.~Policastro, M.~Porrati and P.~van Nieuwenhuizen,  
JHEP {\bf 10} (2002) 054, 
[hep-th/0112162]; 
P.~A.~Grassi, G.~Policastro, and P.~van Nieuwenhuizen,  
JHEP {\bf 11} (2002) 004, 
[hep-th/0202123]; 
P.~A.~Grassi, G.~Policastro and P.~van Nieuwenhuizen, 
Adv.\ Theor.\ Math.\ Phys.\  {\bf 7}, 499 (2003)
[hep-th/0206216].
} 

\lref\asymI{
M.~Bianchi, G.~Pradisi and A.~Sagnotti,
Nucl.\ Phys.\ B {\bf 376}, 365 (1992).
}

\lref\asymII{
P.~Di Vecchia, M.~Frau, I.~Pesando, S.~Sciuto, A.~Lerda and R.~Russo,
Nucl.\ Phys.\ B {\bf 507}, 259 (1997)
[hep-th/9707068]; 
M.~Billo, P.~Di Vecchia, M.~Frau, A.~Lerda, I.~Pesando, R.~Russo and S.~Sciuto,
Nucl.\ Phys.\ B {\bf 526}, 199 (1998)
[hep-th/9802088].}

\lref\asymIII{
N.~Berkovits,
Phys.\ Rev.\ Lett.\  {\bf 79}, 1813 (1997)
[hep-th/9706024];
N.~Berkovits,
Nucl.\ Phys.\ B {\bf 507}, 731 (1997)
[hep-th/9704109]; 
N.~Berkovits and B.~Zwiebach,
Nucl.\ Phys.\ B {\bf 523}, 311 (1998) [hep-th/9711087]; 
}

\lref\pvn{
J.~H.~Schwarz and P.~Van Nieuwenhuizen,
Lett.\ Nuovo Cim.\  {\bf 34}, 21 (1982)\semi
P. van Niewenhuizen, private communication. 
}

\lref\spincoho{
M.~Cederwall, B.~E.~W.~Nilsson and D.~Tsimpis,
JHEP {\bf 0202}, 009 (2002)
[hep-th/0110069]; 
M.~Cederwall, B.~E.~W.~Nilsson and D.~Tsimpis,
JHEP {\bf 0211}, 023 (2002)
[hep-th/0205165].
}

\lref\ChandiaHN{
O.~Chandia and B.~C.~Vallilo,
hep-th/0401226.
}

\lref\GrassiNZ{
P.~A.~Grassi, G.~Policastro and P.~van Nieuwenhuizen,
hep-th/0402122.
} 

\lref\BerkovitsQX{
N.~Berkovits and O.~Chandia,
JHEP {\bf 0208}, 040 (2002)
[hep-th/0204121].
}

\lref\noncom{ J.~de Boer, P.~A.~Grassi and P.~van Nieuwenhuizen,
Phys. Lett.B. [hep-th/0302078]; H.~Ooguri and C.~Vafa, 
[hep-th/0302109]; H.~Ooguri and C.~Vafa, 
hep-th/0303063;
N.~Seiberg,
JHEP 0306 (2003) 010 [hep-th/0305248].
}
\lref\NCsuperspace{S. Ferrara, M.A. Lledo,
JHEP 0005:008,2000 [ hep-th/0002084];
D.~ Klemm, S. ~Penati and L.~ Tamassia,
Class.Quant.Grav.20:2905-2916,2003 [hep-th/0104190]. }
\lref\NACappl{
S.~Ferrara, M.~A. Lledo, and O.~Macia,
JHEP {\bf 0309}, 068 (2003), hep-th/0307039;
M.~Hatsuda, S.~Iso, and H.~Umetsu,
hep-th/0306251;
J.-H. Park,
JHEP {\bf 0309}, 046 (2003), hep-th/0307060;
R.~Britto, B.~Feng, and S.-J. Rey,
JHEP {\bf 0307}, 067 (2003), hep-th/0306215;
R.~Britto, B.~Feng, and S.-J. Rey,
JHEP {\bf 0308}, 001 (2003), hep-th/0307091;
M.~T. Grisaru, S.~Penati, and A.~Romagnoni,
JHEP {\bf 0308}, 003 (2003), hep-th/0307099;
R.~Britto and B.~Feng,
hep-th/0307165;
A.~Romagnoni,
JHEP {\bf 0310}, 016 (2003), hep-th/0307209;
O.~Lunin and S.-J. Rey,
JHEP {\bf 0309}, 045 (2003), hep-th/0307275;
D.~Berenstein and S.-J. Rey,
hep-th/0308049;
R.~Abbaspur,
hep-th/0308050;
A.~Imaanpur,
JHEP {\bf 0309}, 077 (2003), hep-th/0308171;
A.~Imaanpur,
hep-th/0311137;
P.~A. Grassi, R.~Ricci, and D.~Robles-Llana,
hep-th/0311155;
R.~Britto, B.~Feng, and S.-J. Lunin, O.~Rey,
hep-th/0311275;
M.~Billo, M.~Frau, I.~Pesando, and A.~Lerda,
hep-th/0402160;
N.~Berkovits and N.~Seiberg,
JHEP {\bf 0307}, 010 (2003), hep-th/0306226;
S.~Terashima and J.-T. Yee,
hep-th/0306237;
T.~Araki, K.~Ito, and A.~Ohtsuka,
hep-th/0307076;
M.~Chaichian and A.~Kobakhidze,
hep-th/0307243;
E.~Ivanov, O.~Lechtenfeld, and B.~Zupnik,
 hep-th/0308012;
S.~Ferrara and E.~Sokatchev,
hep-th/0308021;
M.~Alishahiha, A.~Ghodsi, and N.~Sadooghi,
hep-th/0309037;
A.~Sako and T.~Suzuki,
hep-th/0309076;
B.~Chandrasekhar and A.~Kumar,
hep-th/0310137;
T.~Araki, K.~Ito, and A.~Ohtsuka,
hep-th/0401012;
C.~Saemann and M.~Wolf,
hep-th/0401147;
T.~Inami and H.~Nakajima,
hep-th/0402137;
A.~Imaanpur and S.~Parvizi,
hep-th/0403174;
D.~Mikulovi\'c,
hep-th/0310065;
D.~Mikulovi\'c,
hep-th/0403290 }
\lref\lie{P. Kosinski, J. Lukierski, P. Maslanka and J. Sobczyk,
J.Phys.A27:6827-6838,1994 [ hep-th/9405076]
}  

\lref\chest{
M.~Chesterman,
JHEP {\bf 0402}, 011 (2004)
[hep-th/0212261]; 
M.~Chesterman,
hep-th/0404021.
}

\lref\Belo{
A.~Belopolsky,
Phys.\ Lett.\ B {\bf 403}, 47 (1997)
[hep-th/9609220]; 
A.~Belopolsky,
hep-th/9703183; 
A.~Belopolsky,
hep-th/9706033.
}

\lref\siegel{
W. Siegel,  and B. Zwiebach, 
Nucl. \ Phys. \ {\bf B}299:206,1988; 
W. Siegel, Phys. Lett. {B} {\bf 142}, 276 (1984), 
Phys. Lett B {\bf 149}, 157 (1984), Phys. Lett. B {\bf 151}, 391 (1985), 
Phys. Lett. B {\bf 149}, 162 (1984), Phys. Lett. B {\bf 151}, 396 (1985). 
} 

\lref\GrassiNZ{
P.~A.~Grassi, G.~Policastro and P.~van Nieuwenhuizen,
hep-th/0402122.
} 

\lref\BerkovitsUA{
N.~Berkovits and M.~M.~Leite,
Phys.\ Lett.\ B {\bf 454}, 38 (1999)
[hep-th/9812153].
}

\lref\dualactions{
N.~Marcus and J.~H.~Schwarz,
Phys.\ Lett.\ B {\bf 115}, 111 (1982);
B.~McClain, F.~Yu and Y.~S.~Wu,
Nucl.\ Phys.\ B {\bf 343}, 689 (1990); 
C.~Wotzasek,
Phys.\ Rev.\ Lett.\  {\bf 66}, 129 (1991);
J.~H.~Schwarz and A.~Sen,
Nucl.\ Phys.\ B {\bf 411}, 35 (1994)
[hep-th/9304154];
P.~Pasti, D.~P.~Sorokin and M.~Tonin,
Phys.\ Rev.\ D {\bf 52}, 4277 (1995)
[hep-th/9506109];
F.~P.~Devecchi and M.~Henneaux,
Phys.\ Rev.\ D {\bf 54}, 1606 (1996)
[hep-th/9603031];
}

\lref\BerkoCSFT{
N.~Berkovits,
Phys.\ Lett.\ B {\bf 388}, 743 (1996)
[hep-th/9607070]. 
}

\lref\BerkoHull{
N.~Berkovits and C.~M.~Hull,
JHEP {\bf 9802}, 012 (1998)
[hep-th/9712007].
}

\lref\BerkovitsUC{
N.~Berkovits,
JHEP {\bf 0209}, 051 (2002)
[hep-th/0201151].
}

\lref\pricom{N. Berkovits, private communication.} 

\Title{
\vbox{\hbox{YITP-SB-04-12} \hbox{IHES/P/04/20} \hbox{FNT/T 2004/04}
}}   
{\vbox{
\centerline{Vertex Operators for Closed Superstrings}
}}  
\vskip -.5cm
\centerline
{
P.~A.~Grassi$^{~a,b,c,}$\foot{pgrassi@insti.physics.sunysb.edu},
and  L.~Tamassia$^{~d,}$\foot{laura.tamassia@pv.infn.it}
} 
\medskip   
\vskip .1cm
\centerline{$^{(a)}$ 
{\it C.N. Yang Institute for Theoretical Physics,} }  
\centerline{\it State University of New York at Stony Brook,   
NY 11794-3840, USA,}  
\medskip
\centerline{$^{(b)}$ {\it Dipartimento di Scienze,
Universit\`a del Piemonte Orientale,}}
\centerline{\it
C.so Borsalino 54, I-15100 Alessandria, ITALY, and}
\medskip
\centerline{$^{(c)}$ {\it 
IHES, Le Bois-Marie, 35, route de Chartres, F-91440 Bures-sur-Yvette, FRANCE}}
\medskip
\centerline{$^{(d)}$ {\it Dipartimento di Fisica Nucleare e Teorica, Universit\`a degli studi di Pavia}}
\centerline{\it and INFN, Sezione di Pavia, via Bassi 6, I-27100 Pavia, ITALY}
\vskip .3cm  

\smallskip
\bigskip

\centerline{\bf Abstract}
\smallskip

We construct an iterative procedure to compute the vertex operators 
of the closed superstring in the covariant formalism 
given a solution of IIA/IIB supergravity. 
The manifest supersymmetry 
allows us to construct vertex operators for any generic background in presence 
of Ramond-Ramond (RR) fields. We extend the procedure to all massive states of open 
and closed superstrings and we identify two new nilpotent charges which are used 
to impose the gauge fixing on the physical states. We solve iteratively the equations of the 
vertex for linear $x$-dependent RR field strengths. This vertex plays a role in studying 
non-constant C-deformations of superspace. Finally, we construct an action for the free 
massless sector of closed strings, and we propose a form for the kinetic term for closed 
string field theory in the pure spinor formalism. 

\bigskip

\Date{\newdate}

\listtoc
\writetoc

\baselineskip 14pt 


\newsec{Introduction}

Motivated by the increasing interest in the covariant 
techniques for computation of the amplitudes 
in string theory, we provide a calculation scheme for vertex operators 
in pure spinor approach string theory in 10 dimensions 
\refs{\berko,\grassi}. 
The covariant methods turned out to be superior in order to derive manifest super-Poincar\'e invariant effective actions and to handle generic backgrounds (for example with RR fields such as $AdS_{5} \times S^{5}$ and 
pp-waves) avoiding, for instance, GSO projections, sums over spin structures and light-cone contact terms.
However, since the amount of symmetries that are manifest in the covariant formulation increases, also the number of auxiliary fields 
increases and a useful technique to compute 
the basic ingredients is needed.  Here we provide such a 
procedure and some applications. 
First of all we give a brief review of the open superstring formalism, 
we explain the main idea and we outline the rest of the paper.  

In the case of the open superstring, the massless sector is described by a 
vertex operator ${\cal V}^{(1)} = 
\l^{\a} A_{\a}$ at ghost number one  where $\l^{\a}$ is 
a pure spinor  
(defined in appendix A) and $A_{\a}(x,\t)$ 
is the spinorial component of the superconnection. 
The superfield $A_{\a}$ is completely 
expressed in terms of the 
gauge field $a_{m}(x)$ and the gluino $\psi^{\a}(x)$, for example as
\eqn\I{
A_{\a}(x,\t) = {1\over 2}(\g^m\t)_\a a_m (x) +{1\over 3} (\g^m\t)_\a (\g_m\t)_\g \psi^\g (x) + {\cal O}(\t^3). 
}
The vertex operator ${\cal V}^{(1)}$ belongs to the cohomology of the 
BRST charge $Q = \int d\sigma \l^{\a} d_{\sigma\a}$, where $d_{\sigma\a}$ 
is defined in app. A, if and only if 
the components of $A_{\a}$ satisfy the linear Maxwell and Dirac equations
\eqn\IA{
\p^{m} (\p_{m} a_{n} - \p_{n} a_{m}) =0 \,, ~~~~~
\g^{m}_{\a\b} \p_{m} \psi^{\b} = 0\,.
}
The contributions ${\cal O}(\t^{3})$ are given in terms of the derivatives 
of $a_{m}$ and $\psi$ and are completely fixed by the equations 
of motion 
\eqn\IB{
D_{(\a} A_{\b)} - \g^{m}_{\a\b} A_{m}= 0
} given in \howe, where 
$A_{m}$ is the vectorial part of the superconnection and  
$D_{\a} = \p_\a+ {1\over 2}(\g^m\t)_\a \p_m$ is the superderivative.  
The lowest components of $A_{\a}$ in \I\ are eliminated by  a gauge fixing condition. 

Even though the computation of all terms in the expansion of $A_{\a}$ seems a 
straightforward procedure, technically it is rather involved. However, 
there exists a powerful technique which simplifies the task. 
The main idea is to choose a suitable gauge fixing such as for instance
\eqn\II{
\t^{\a} A_{\a}(x, \t) = 0\,,
}
which reduces the independent components in the superfield $A_{\a}$.
This choice\foot{The following 
gauge condition has a counterpart in bosonic string theory: $x^{m} A_{m}(x) =0$.
This fixes the gauge invariance under $\delta A_{m} = \p_{m} \omega(x)$ and 
it coincides with the Lorentz gauge in momentum space $\p_{p_{\mu}} \tilde A_{m} =0$. 
The gauge fixing yields the equation $( 1 + x^{n} \p_{n})  A_{m} = x^{n} F_{mn}$
which can be solved directly by inverting $(1 + x^{n} \p_{n})$ and obtaining 
$A_{m} = \int^{x} d^{26}y [(1 + y^{p} \p_{p})^{-1} (y^{n} F_{mn}(y) ]$.
} fixes part of the super-gauge transformation 
$\delta A_{\a} = D_{\a} \Omega$, where $\Omega$ is a scalar 
superfield with ghost number zero. To reach the gauge \II, we have 
to impose $\t^{\a}(A_{\a} + \delta A_{\a}) = 0$, which implies 
that $\t^{\a} D_{\a} \Omega = - \t^{\a}A_{\a}$. Expanding 
$\Omega$ as $\Omega = \sum_{n \geq 0} 
\Omega_{[\a_{1} \dots \a_{n}]} \t^{\a_{1}} \dots \t^{\a_{n}}$, all components with 
$n\geq 1$ are fixed except the lowest component $\Omega_{0}$, which corresponds to the usual bosonic gauge transformation 
of Maxwell theory. 
 
Acting with $D_{\a}$ on \II\ and 
using the equations of motion \IB, one gets the recursive relations 
\eqn\III{\eqalign{
&(1 + {\bf D}) A_{\a} =  (\g^{m} \t)_\a A_{m}\cr
&{\bf D}A_m=(\g_m\t)_\g W^\g \cr
&{\bf D}W^\a=-{1\over 4}(\g^{mn}\t)^\a F_{mn}\cr
&{\bf D}F_{mn}=-(\g_{[m}\t)_\g \p_{n]} W^\g
}}
where ${\bf D} \equiv \t^{\a} \p_{\a}$. So, given the zero-order component of $A_{m}$, we can compute the order-$\t$ component of $A_{\a}$. 
The same can be done for $A_{m}$, the spinorial field strength $W^{\a}$ and 
the bosonic curvature $F_{mn} = \p_{[m} A_{n]}$ making use of the other three equations. 
This renders the  
task of computing all components of $A_{\a}$ in terms of initial data 
$A_{m}(x) = a_{m}(x) + {\cal O}(\t)$ and 
$W^{\a}(x)=\psi^{\a}(x) + {\cal O}(\t)$ a purely algebraic problem
(\har\ and \OoguriPS). 
Moreover, 
one is able to compute all components of the superfields 
appearing in the (descent) ghost-number-zero vertex operator ${\cal V}^{(0)}_{\sigma}$ 
\eqn\IV{
{\cal V}^{(0)}_{\sigma} = \p_{\sigma} \t^{\a} A_{\a} + \Pi^{m}_{\sigma} A_{m} 
+ d_{\sigma\a} W^{\a} + \demi\, N^{mn}_{\sigma} F_{mn}\,, 
}
which satisfies the descent equation 
$[Q, {\cal V}^{(0)}_{\sigma}] = \p_{\sigma} {\cal V}^{(1)}$. 
Here $\sigma$ is the boundary worldsheet coordinate and 
$N^{mn}_{\sigma} = {1\over 2} w_{\sigma} \g^{mn} \l$ is the pure spinor part of the Lorentz current. The operators 
$\Pi^{m}_{\sigma}$ and $d_{\sigma\a}$ are the supersymmetric line 
element and the fermionic constraint of the Green-Schwarz superstring \GS, 
respectively. 	

In the present paper, we apply the same technique to 
IIA/IIB supergravity. We start from the vertex operators for closed 
superstrings, we derive the complete set of equations 
from the BRST cohomology, we define all curvatures and 
gauge transformations. Then, we impose a set of gauge fixing 
conditions to remove the lowest components of the superfields and 
we derive an iterative procedure to compute all components. We show 
that we need a further gauge fixing to fix the reducible gauge symmetries 
and we show that all chosen gauges can indeed be reached. 

The procedure for closed strings is original by itself, but, more importantly, 
the present analysis leads to a generalization of \II\ to all 
vertex operators, associated to both massless and massive states. Indeed, we will show that the 
gauge fixing \II\ can be written in terms of a new nilpotent 
charge ${\cal K}$ (with negative ghost number) as follows
\eqn\IVA{
\{ {\cal K}, {\cal V}^{(1)} \}= 0 \,.
}
This imitates the Siegel gauge in string field theory. When restricted to massless states, this generalized gauge fixing condition reduces to the gauge fixing \II\ for open superstrings and to the corresponding gauge fixing for closed strings discussed in the text. 
When applied to massive states, \IVA\ also leads to a suitable gauge fixing. In the paper, we explicitly derive the gauge conditions for the first massive state for the open superstring. 
Again, \IVA\ fixes all auxiliary fields in terms of the physical on-shell data and 
eliminates the lowest components. 

As an application of the computation technique we calculate 
the vertex operator of linear $x$-dependent RR field strength. This amounts 
to expanding the superfields for closed strings up to fourth order in 
powers of both $\t$ and $\hat\t$ (where the hatted quantities refer to the right-moving sector 
of the theory). This new vertex operator is the starting point for studying $x$-dependent 
$C$-deformations which might give new insight in the superspace 
structure of  supergravity  \pvn. 

As pointed out in \refs{\asymI,\asymII,\asymIII}, in 
the RNS framework \fms\ vertex operators in the asymmetric picture 
are very useful to study the dynamics of D-branes. In fact, 
in the asymmetric picture the vertex operator is not only expressed in terms of the RR field strength, but it is also parametrized by the RR potential. The analysis in \asymI\ shows that 
the off-shell vertex operators directly couple the RR potentials to the worldvolume of the
D-brane. We present a new method to relax the superspace constraints by adding 
new auxiliary fields and constructing the corresponding vertex operators.
\foot{The construction of vertices with RR potentials in covariant formulation 
has been also discussed in 
\lref\CornalbaCU{
L.~Cornalba, M.~S.~Costa and R.~Schiappa,
hep-th/0209164.
} \CornalbaCU. There the authors consider constant RR potentials. } 
We show that it leads to a deformation of the superspace constraints 
(that, in 10 dimensions, force the physical fields to be on-shell), generalizing the method presented in 
\spincoho\ for N=1 D=10 SYM theory \SYM. The gauge transformations 
are studied and the vertex operators are expressed in terms of the RR potential. 

In the present paper we only consider deformations (vertex operators) 
at first order in the coupling constant, neglecting the backreaction 
of background fields. For the complete sigma model for 
open, closed and heterotic strings see for example 
\BerkovitsUE\ and the one-loop computations \ChandiaHN. 

The linearized form of supergravity equations written in terms of the BRST charges of the pure spinor 
sigma model gives us the framework to analyze some aspects of 
closed string field theory action. As is well-known, the action for closed string field theory 
has to take into account the presence of selfdual forms (for example the 
five form in type IIB supergravity). This can be done either by breaking explicitly 
the Lorentz invariance, or by admitting an infinite number of fields 
in the action \dualactions. We 
show that this action can be indeed constructed mimicking the bosonic 
closed string field theory action 
discussed in \zwie\ (and in the  references therein). We show that we can 
easily account for new fields which nevertheless do not propagate, and 
we check that the action has the correct symmetries leading to the complete BV 
action for type IIA/IIB supergravity. 

In section 2, we derive the complete set of descent equations for 
vertex operators of closed superstrings and their gauge transformations. 
Moreover, we derive the consistency conditions for the gauge parameters. 
In section 3, we compute the complete set of equations in superspace and the equations 
of motion for the physical fields. 
In section 4, we construct the gauge transformations in terms 
of ``physical'' gauge parameters and we provide the gauge fixing conditions.
Section 5 deals with the  iterative 
procedure to extract the relations among supergravity fields and 
auxiliary fields. In section 6, we generalize the gauge conditions to 
massive states and we study the first massive state. 
The form of the vertex for non constant RR fields is given 
in section 7. Section 8 deals 
with an off-shell formulation of closed string vertex operators.  
Finally, in section 9 we make a proposal for the kinetic terms of a string field theory action for closed 
strings. 
In appendix A, we give some basic formulas for pure spinor superstring theory. In appendix B, we give the expansion of the auxiliary superfields up to order two in both fermionic coordinates, obtained by solving the iterative equations presented in section 5. In appendix C part of the superfields needed in the computation of the vertex for a linear $x$-dependent RR field strength are given. 

  
\newsec{Vertex Operators for Closed Superstrings} 

  To compute correlation functions we need not only the ghost number 
  $(1,1)$ local vertex   ${\cal V}^{(1,1)}$ , but we also 
  need the integrated vertex operators $\int dz~ {\cal
  V}_z^{(0,1)}$, $\int d\bar z~ {\cal V}_{\bar z}^{(1,0)}$, and $\int dz \wedge d\bar z~ {\cal V}_{z\bar z}^{(0,0)}$. They satisfy the descent equations which we are 
  going to discuss.  
  
  Introducing the notation 
  ${\cal O}^{(a,b)}_{c,d}$ for local vertex operators with ghost number $a (b)$ 
  in the left (right) sector and (anti)holomorphic indices $c (d)$, 
  we identify 
  \eqn\deA{
  {\cal O}^{(1,1)}_{0,0} = {\cal V}^{(1,1)}\,, ~~~~~
  {\cal O}^{(0,1)}_{1,0} = {\cal V}^{(0,1)}_z dz\,, ~~~~~
  {\cal O}^{(1,0)}_{0,1} = {\cal V}^{(1,0)}_{\bar z} d \bar z\,, ~~~~~
  {\cal O}^{(0,0)}_{1,1} = {\cal V}^{(0,0)}_{z\bar z} dz\wedge d\bar z\,.
  }
  The descent equations read\foot{We use the square backets to denote both 
  commutation and anti-commutation relations. The difference is established by 
  the nature of the operators involved in the relations. 
  } 
  \eqn\deB{
  [Q_L, {\cal O}^{(a,b)}_{c,d}] = \p \, {\cal O}^{(a+1,b)}_{c-1,d}\,, 
  ~~~~~
  [Q_R, {\cal O}^{(a,b)}_{c,d}] =\bar \p \, {\cal O}^{(a,b+1)}_{c,d-1}\,,
  }
  where $\p = dz \p_z$ and $\bar \p = d\bar z \p_{\bar z}$ 
  are the holomorphic and antiholomorphic differentials. $Q_L$ and 
  $Q_R$ are the BRST charges for holomorphic and antiholomorphic 
  sectors, satisfying the anticommutation relation $[Q_L,Q_R]=0$ 
  (for their explicit form in terms of sigma model fields, see app. A). 
  More explicitly, at the first level we have 
  \eqn\deC{
  [Q_L, {\cal V}^{(1,1)}] = 0 \,,
  ~~~~~
  [Q_R, {\cal V}^{(1,1)}] = 0\,,
  }
  while at the next level we get 
  \eqn\deCC{
  [Q_L, {\cal V}^{(0,1)}_z] = \p_z {\cal V}^{(1,1)}\,, 
  ~~~~~
  [Q_R, {\cal V}^{(0,1)}_z] = 0\,,
  }
  $$
  [Q_R, {\cal V}^{(1,0)}_{\bar z}] = \p_{\bar z} {\cal V}^{(1,1)}\,, 
  ~~~~~
  [Q_L, {\cal V}^{(1,0)}_{\bar z}] = 0\,,
  $$
  and, finally, 
  \eqn\deD{
  [Q_L, {\cal V}^{(0,0)}_{z \bar z}] = \p_z  {\cal V}^{(1,0)}_{\bar z} \,, 
  ~~~~~
  [Q_R, {\cal V}^{(0,0)}_{z \bar z}] = - \p_{\bar z} {\cal V}^{(0,1)}_{z} \,.
  }
The vertex operators ${\cal O}^{(a,b)}_{c,d}$ are expanded 
in powers of ghost fields $\l^\a$ and $\hat\l^{\hat \a}$ or in powers 
of the supersymmetric holomorphic and antiholomorphic 1-forms
 \eqn\vectors{\eqalign{{\bf X}_z =  \left( \p_z\theta^\a,~ \Pi^m_z,~ d_{z\a},~ 
{1\over 2} N^{mn}_z \right)\,, ~~~~~~~ 
{\bf \hat X}_{\bar z} = \left( \pa_{\bar z}\hat\theta^{\hat\b}~,~
\hat\Pi^p_{\bar z}~,~ \hat d_{\bar z\hat \b}~,~ {1\over 2} \hat N^{pq}_{\bar z}\right)\,.}} 
The explicit expression of these 1-form operators in terms of sigma model fields is given in appendix A. 
The coefficients are superfields of the coordinates $x^m$, $\t^\a$ and 
$\hat\t^{\hat \a}$. A further relation is obtained by acting from the left on the first equation of \deD\ with $Q_R$ or on the second with $Q_L$. Using eqs. 
\deCC\ and the commutation relations (A.3), 
one obtains 
\eqn\deE{
[Q_R,[Q_L, {\cal V}^{(0,0)}_{z \bar z}]] = \p_z  \p_{\bar z} {\cal V}^{(1,1)}\,,
} 
which turns out to be useful for the explicit computations in sec. 3. 

Equations \deC, \deCC, and \deD\ 
are invariant under the gauge transformations given by 
$$
\d {\cal V}^{(1,1)}= [Q_L, \L^{(0,1)}]+ [Q_R, \L^{(1,0)}] 
$$ 
\eqn\gaugegen{
\d {\cal V}_{\bar z}^{(1,0)} = [Q_L,  \tau _{\bar z}^{(0,0)}] + \p_{\bar z} \L^{(1,0)}\,, 
~~~~~~~~~
\d {\cal V}_z^{(0,1)}= [Q_R, \tau_z ^{(0,0)}] + \p_z \L^{(0,1)} 
}
$$
\d {\cal V}_{z \bar z}^{(0,0)}=\p_z \tau ^{(0,0)}_{\bar z} - 
\p_{\bar z} \tau_z^{(0,0)}
$$
where the zero forms $\L^{(0,1)}$ and $ \L^{(1,0)}$ have ghost number $(1,0)$ and 
$(0,1)$ and are proportional to  $\l^\a$ and $\hat\l^{\hat \a}$, 
and the coefficients are superfields. 
The holomorphic and antiholomorphic 
1-forms $\tau_z ^{(0,0)}$ and $\tau ^{(0,0)}_{\bar z}$ are 
to be expanded in terms of powers of the 1-forms ${\bf X}_z$ and ${\bf \hat X}_{\bar z}$ given in \vectors\ and their coefficients are again superfields. 

In addition, the gauge parameters $\L^{(0,1)}$, $\L^{(1,0)}$, $\tau_z ^{(0,0)}$ and $\tau ^{(0,0)}_{\bar z}$ must satisfy the following consistency conditions
\eqn\consI{\eqalign{ 
[Q_L, \L^{(1,0)}]=0 \quad\quad\quad 
[Q_R, \L^{(0,1)}]=0\,,
}}
and
\eqn\consII{\eqalign{
[Q_L, \tau_z^{(0,0)}] + \p_z \L^{(1,0)}=0 \quad\quad\quad
[Q_R, \tau_{\bar z}^{(0,0)}] + \p_{\bar z} \L^{(0,1)}=0\,.
}}
These equations resemble the descent equations 
for the  open string vertex operator ${\cal V}^{(1)}= \l^\a A_\a$, 
but in that case there are boundary conditions 
for the fermionic fields: $\t^\a(z) = \hat\t^{\hat \a}(z)$ at $z = \bar z$.  

Equations \consI\ and \consII\ are further invariant under the gauge transformations 
\eqn\constrans{
\delta \Lambda^{(1,0)} = [Q_{L}, \Upsilon^{(0,0)}]\,, ~~~~~
\delta \Lambda^{(0,1)} = [Q_{R}, \hat\Upsilon^{(0,0)}]\,,
}
$$
\delta \tau^{(0,0)}_{z} =- \p_{z} \Upsilon^{(0,0)}\,, ~~~~~
\delta \tau^{(0,0)}_{\bar z} =- \p_{\bar z} \hat\Upsilon^{(0,0)}\,. 
$$
where $\Upsilon^{(0,0)}$ and $\hat\Upsilon^{(0,0)}$ are generic superfields. However, 
consistency with \gaugegen\ imposes $\Upsilon^{(0,0)}= \hat\Upsilon^{(0,0)}$. 
The superfield $\Upsilon^{(0,0)}$ will be useful to define a suitable gauge fixing  procedure 
and to take into account the reducible gauge symmetry of the 
NS-NS two form of 10-dimensional supergravity. 

To derive equations \deCC\ we can view the vertex operators 
${\cal V}^{(0,1)}_z$ and 
${\cal V}^{(1,0)}_{\bar z}$ 
as deformations of the BRST charges (A.2)
\eqn\defBRST{
Q_L \rightarrow Q_L + \oint d\bar z \, {\cal V}^{(1,0)}_{\bar z}\,, 
\quad\quad\quad
Q_R \rightarrow Q_R + \oint d z \, {\cal V}^{(0,1)}_{z}\,, 
}
and the vertex operator ${\cal V}^{(0,0)}_{z \bar z}$ as
the deformation of the action
\eqn\defAction{
S \rightarrow S + \int dz d\bar z \, {\cal V}^{(0,0)}_{z \bar z}\,. 
}
Eqs. \deC\ are derived by requiring the nilpotency of the new charges and their anticommutation relations. 

In terms of the vertex operators ${\cal O}^{(a,b)}_{c,d}$, the amplitudes on the 
sphere are defined in \BerkovitsPH\ as
\eqn\ampl{
{\cal A}_{n+3} = 
\Big\langle\Big\langle  
{\cal V}^{(1,1)}(z_{1},\bar z_{1}) {\cal V}^{(1,1)}(z_{2},\bar z_{2}) {\cal V}^{(1,1)}(z_{3},\bar z_{3})
\prod_{n} \int dz d\bar z {\cal V}^{(0,0)}
\Big\rangle\Big\rangle
} 
 where the three unintegrated vertex operators are needed to fix the 
 $SL(2,{\bf C})$ invariance on the sphere. An unintegrated vertex 
 ${\cal V}^{(1,1)}(z_{1},\bar z_{1})$ can be replaced by a product of $(1,0)$ and 
 $(0,1)$ vertices $\oint dz {\cal V}_{z}^{(0,1)} \oint d\bar z {\cal V}_{\bar z}^{(1,0)}$ 
 which has the same total ghost number and the same total 
 conformal spin as the original vertex ${\cal V}^{(1,1)}$. In \BerkovitsPH\ 
 supersymmetry and gauge invariance were proven 
 under the assumption that the prescription for the zero modes is the following
 \eqn\zeromodi{
 \langle\langle  {\cal V}^{(3,3)} \rangle\rangle = 1
}
where 
$$
 {\cal V}^{(3,3)} =  
 (\l_{0}\g^{m}\t_{0} 
\l_{0}\g^{n}\t_{0} \l_{0}\g^{p}\t_{0} \t_{0} \g_{mnp}\t_{0}) 
(\hat\l_{0}\g^{m} \hat\t_{0} 
\hat\l_{0}\g^{n}\hat\t_{0} \hat\l_{0}\g^{p}\hat\t_{0} \hat\t_{0} \g_{mnp}\hat\t_{0}) \,. 
$$

  
  \newsec{Equations of Motion} 
   
In the present section we derive the equations of motion for the 
massless background 
fields in superspace from the BRST cohomology of the superstring. 
Let us start from the simplest equations \deC\ for the 
vertex ${\cal V}^{(1,1)}$ whose general expression is 
  \eqn\verB{ 
  {\cal V}^{(1,1)} = \l^\a A_{\a\hat\b}\hat\l^{\hat\b} \,. 
  } 
 The superfield $A_{\a \hat\b}(x,\t,\hat\t)$ 
  satisfies the equations of motion  \howe\
  \eqn\verC{ 
  \g^{\a\b}_{mnopq} D_\a A_{\b \hat\b} =0\,, ~~~~~~ 
  \g^{\hat\a\hat\b}_{mnopq} D_{\hat\a} A_{\a \hat\b} =0\,, ~~~~~~ 
  } 
  where $\g^{\a\b}_{mnopq}$ is the antisymmetrized product of 
  five gamma matrices. The pure spinor conditions imply that only the 
 5-form parts of the $D_\a A_{\b \hat\b}$ and $D_{\hat\a} A_{\a \hat\b}$ 
 are indeed constrained \berko. 
 By using Bianchi identities, one can show that 
 they yield the type IIA/IIB supergravity equations of motion at the linearized level. 
 All auxiliary fields present in the superfield $A_{\a\hat \b}$ are fixed by 
 eqs.\verC. 
 
 As outlined before, one can use  
 different types of vertices to simplify the computations. 
 Integrated vertices are written 
 in terms of a huge number of different superfields, whose components 
 are completely fixed by the equations of 
 motion. As a result, these vertices are quite complicated espressions.

The set of superfields needed to compute 
 ${\cal V}^{(0,0)}, \dots, {\cal V}^{(1,1)}$
 can be grouped into the following matrix
 \eqn\vertmatrix{ 
{\bf A } = 
 \left[ \matrix{A_{\a\hat\b} & A_{\a p} & E_{\a}^{~~\hat\b}& 
 \Omega_{\a, pq}\cr
A_{m\hat\b} & A_{mp} & E_m^{~~\hat\b} & 
\Omega_{m, pq}\cr
 E^{\a}_{~~\hat\b} & E^\a_{~~p} & P^{\a\hat\b} & 
 {C}^\a_{~~pq}\cr
 \Omega_{mn, \hat\b} & \Omega_{mn,p} & 
 C_{mn}^{~~~~\hat\b} & S_{mn,pq}}\right]
 }
The first components of $A_{mp}$, 
$E_m^{~~\hat\b}$, $E^\a_{~~p}$ and $P^{\a\hat\b}$ 
are identified with the  supergravity fields as follows
\eqn\sugra{
A_{mp}= g_{mp} + b_{mp} + \eta_{mp} \phi + {\cal O}(\t, \hat\t)\,, 
}
$$
E_m^{~~\hat\b} = \psi_m^{~~\hat \b} +  {\cal O}(\t, \hat\t)\,,  ~~~~~~~~~E^\a_{~~p}  =  \psi_{~~p}^\a + {\cal O}(\t, \hat\t)\,, 
$$
$$ P^{\a\hat\b} = f^{\a\hat\b} + {\cal O}(\t, \hat\t)\,.
$$
The fields $g_{mn}$, $b_{mn}$, $\phi$, 
 $\psi_{~~p}^\a$, $\psi_m^{~~\hat \b}$ and $f^{\a \hat \b}$ are 
 the graviton, the NS-NS two-form, the dilaton, the two gravitinos (the gamma-traceless 
 part of $\psi_{~~p}^\a$, $\psi_m^{~~\hat \b}$), the two dilatinos (the 
 gamma-trace part of $\psi_{~~p}^\a$, $\psi_m^{~~\hat \b}$) and the
 RR field strengths. IIA and IIB differ in the chirality 
 of the two spinorial indices $\a$ and $\hat \a$. This 
 changes the type of RR fields present in the spectrum. The first 
 components of the superfields $\Omega_{m,pq}$ ($\Omega_{mn,p}$), $C_{mn}^{~~~~\b}$  ($C^\a_{~~ pq}$) and $S_{mn,pq}$ are identified with the linearized 
 gravitational connection $\Gamma_{rs}^t$, the curvature of 
 the gravitinos and the linearized Riemann tensor,  respectively.
 The remaining superfields are the spinorial partners of the above superfields. 
 In ten dimensions, the superspace constraints together with the Bianchi identities imply the supergravity equations of motion 
 \lref\sugra{
L.~Brink and P.~S.~Howe,
Phys.\ Lett.\ B {\bf 88}, 81 (1979).
P.~S.~Howe,
Nucl.\ Phys.\ B {\bf 199}, 309 (1982).
P.~S.~Howe and P.~C.~West,
Nucl.\ Phys.\ B {\bf 238}, 181 (1984);
 J. L. Carr, J.S.James Gates,Jr., R. N. Oerter
Phys.Lett. B {\bf 189}, 68 (1987). 
} \sugra. 
 Those constraints are 
 given in terms of the spinorial components $A_{\a\hat\b}$, $A_{\a p}$, $E_{\a}^{~~\hat\b}$ 
 and $\Omega_{\a, pq}$. The structure of superspace formulation 
 of type IIA and IIB supergravity in the present framework is also 
 discussed in \BerkovitsUE. 

Given the vectors ${\bf X}_z$ and ${\bf \hat X}_{\bar z}$ (see \vectors)
we can explicitly write the vertex operator ${\cal V}^{(0,0)}_{z \bar z} ={\bf X}^T_z {\bf A} {\bf \hat X}_{\bar z} $ 
as
\eqn\vert{\eqalign{ 
{\cal V}^{(0,0)}_{z \bar z} 
& 
= \pa_z\theta^\a~ A_{\a\hat\b}~ \pa_{\bar z}\hat\theta^{\hat\b} 
+ \pa_z \theta^\a~ A_{\a p}~ \hat\Pi^p_{\bar z} 
+ \Pi^m_z~ A_{m \hat\b}~ \pa_{\bar z}\hat\theta^{\hat\b}
+ \Pi^m_z~ A_{mp}~ \hat\Pi^p_{\bar z}  \cr
& 
+ d_{z\a}~ E^\a_{~~\hat\b}~ \pa_{\bar z}\hat\theta^{\hat\b}~ 
+ d_{z\a}~ E^\a_{~~p}~ \hat\Pi^p_{\bar z} 
+ \pa_z\theta^\a~ E_\a^{~~\hat\b}~\hat d_{\bar z\hat\b}
+ \Pi^m_z~ E_m^{~~\hat\b}~\hat d_{\bar z\hat\b} 
+ d_{z\a}~ P^{\a\hat\b}~ \hat d_{\bar z\hat\b}\cr
& 
+{1\over 2}~ N^{mn}_z~ \Omega_{mn,\hat\b}~ \pa_{\bar z}\hat\theta^{\hat\b} 
+{1\over 2}~ N^{mn}_z~ \Omega_{mn,p}~ \hat\Pi^p_{\bar z} 
+{1\over 2}~ \pa_z\theta^\a~ \Omega_{\a,pq} \hat N^{pq}_{\bar z} 
+{1\over 2}~ \Pi^m_z~  \Omega_{m,pq} \hat N^{pq}_{\bar z} \cr
& 
+{1\over 2}~ N^{mn}_z~ C_{mn}^{\quad\hat\b}~\hat d_{\bar z\hat\b} 
+{1\over 2}~ d_{z\a}~ C^{\a}_{\quad pq}~ \hat N^{pq}_{\bar z} 
+{1\over 4}~ N^{mn}_z~ S_{mn,pq}~ \hat N^{pq}_{\bar z}
}}
\vskip 12pt
From equations \deC, \deCC, \deD\ and \deE\ 
in the previous section we derive the complete set of 
equations for the background fields
\vskip 12pt
\eqn\eom{\eqalign{
&^{\left({1\over 2},{1\over 2},{1\over 2}\right)}
~~~~D_\a A_{\b\hat\g} + D_\b A_{\a\hat\g} - \g^m_{\a\b} A_{m\hat\g}=0 ~~~~~~~~~ \hat D_{\hat\a} A_{\b\hat\g} + \hat D_{\hat\g} 
A_{\b\hat\a} - \g^m_{\hat\a\hat\g} A_{\b m}=0\cr
&^{\left({1\over 2},{1\over 2}, 1\right)} ~~~~ D_\a A_{m\hat\b} -\pa_m A_{\a\hat\b}-\g_{m\a\g}E^{\g}_{~~\hat\b}=0 ~~~~~~~ \hat D_{\hat\a} A_{\b p} - \pa_p A_{\b\hat\a} - \g_{p\hat\a \hat\g} E_\b^{~~\hat\g}=0\cr
&^{\left({1\over 2},{1\over 2}, 1\right)} ~~~~ D_\a A_{\b p} + D_\b A_{\a p} -\g^m_{\a\b} A_{mp}=0 ~~~~~~~~~~ \hat D_{\hat\a}A_{m\hat\b}
+ \hat D_{\hat\b} A_{m\hat\a} + \g^p_{\hat\a\hat\b}A_{mp}=0\cr
&^{\left({1\over 2},{1\over 2}, {3\over 2}\right)} ~~~~ D_\a E^{\b}_{~~\hat\g} - {1\over 4} (\g^{mn})_\a^{~~\b} \Omega_{mn,\hat\g}=0 ~~~~~~~~~~~ \hat D_{\hat\a} E_\b^{~~\hat\g} -{1\over 4} (\g^{pq})_{\hat\a}^{~~\hat\g} \Omega_{\b, pq}=0\cr
&^{\left({1\over 2},{1\over 2}, {3\over 2}\right)} ~~~~ D_\a E_\b^{~~\hat\g} + D_\b E_\a^{~~\hat\g} - \g^m_{~~\a\b} E_m^{~~\hat\g}=0 ~~~~~~ \hat D_{\hat\a} E^\b_{~~\hat\g} + \hat D_{\hat\g} E^\b_{~~\hat\a} - \g^p_{~~\hat\a\hat\g}E^\b_{~~p}=0\cr
&^{\left({1\over 2},1, 1\right)} ~~~~ D_\a A_{mp} -\pa_m A_{\a p} - \g_{m\a\g} E^\g_{~~p}=0 ~~~~~~~~~ \hat D_{\hat\a}A_{mp}+ \pa_p A_{m\hat\a} + \g_{p\hat\a\hat\b}E_m^{~~\hat\b}=0\cr
&^{\left({1\over 2},{3\over 2}, 1\right)} ~~~~ D_\a E^\b_{~~p} - {1\over 4} (\g^{mn})_\a^{~~\b}\Omega_{mn,p}=0 ~~~~~~~~~~~~ \hat D_{\hat\a} E_m^{~~\hat\b}+{1\over 4} \Omega_{m,pq}(\g^{pq})_{\hat\a}^{~~\hat\b}=0\cr
&^{\left({1\over 2},{3\over 2}, 1\right)} ~~~~ D_\a E_m^{~~\hat\b} -\pa_m E_\a^{~~\hat\b} - \g_{m\a\g}P^{\g\hat\b}=0 ~~~~~~~~ \hat D_{\hat\a} E^\b_{~~p} - \pa_p E^\b_{~~\hat\a} - \g_{p\hat\a\hat\g}P^{\b\hat\g}=0\cr
&^{\left({1\over 2},{1\over 2}, 2\right)} ~~~~ D_\a \Omega_{\b, pq} + D_\b \Omega_{\a, pq} - \g^m_{~~\a\b}\Omega_{m,pq}=0 ~~~ \hat D_{\hat\a} \Omega_{mn,\hat\b} + \hat D_{\hat\b} \Omega_{mn,\hat\a}+ \g^p_{\hat\a\hat\b}\Omega_{mn,p}=0\cr
&^{\left({1\over 2},{3\over 2}, {3\over 2}\right)} ~~~~ D_\a P^{\b\hat\g} -{1\over 4} (\g^{mn})_\a^{~~\b}C_{mn}^{~~~~\hat\g}=0 ~~~~~~~~~~~~ \hat D_{\hat\a} P^{\b\hat\g} - {1\over 4} (\g^{pq})_{\hat\a}^{~~\hat\g}C^{\b}_{~~pq} =0\cr
&^{\left({1\over 2},1, 2\right)} ~~~~ D_\a \Omega_{m,pq} -\pa_m \Omega_{\a, pq}- \g_{m\a\g}C^{\g}_{~~ pq}=0
~~~~~ \hat D_{\hat\a}\Omega_{mn,p} + \pa_p \Omega_{mn,\hat\a} + \g_{p\hat\a\hat\b}C_{mn}^{~~~~\hat\b}=0\cr
&^{\left({1\over 2},{3\over 2}, 2\right)} ~~~~D_\a C^{\b}_{~~pq} -{1\over 4} (\g^{mn})_\a^{~~\b}S_{mn,pq}=0 ~~~~~~~~~~~ \hat D_{\hat\a} C_{mn}^{~~~~\hat\b} + {1\over 4} (\g^{pq})_{\hat\a}^{~~\hat\b}S_{mnpq}=0
}}
\vskip 12pt
where the labels $(a,b,c)$ denote the scaling dimensions of the generators of  
the extended super-Poincar\'e algebra 
(\GreenNN\ and \csm) 
which the equations belong to. 

Moreover, one obtains the following eight equations, which do not provide further information, since they are implied by \eom\ and pure spinor conditions given in app. A
\vskip 12pt
\eqn\trivialeq{\eqalign{&N^{mn}\l^\g D_\g \Omega_{mn,\hat\b}=0~~~~~~~~~~ \hat\l^{\hat\g}\hat D_{\hat\g}\Omega_{\a, mn}\hat N^{mn}=0\cr
 &N^{mn} \l^\g D_\g \Omega_{mn,p}=0~~~~~~~~~~ \hat\l^{\hat\g}\hat D_{\hat\g}\Omega_{m,pq}\hat N^{pq}=0\cr
 &N^{mn} \l^\g D_\g C_{mn}^{~~~~\hat\b}=0~~~~~~~~~~\hat\l^{\hat\g}\hat D_{\hat\g}C^{\a}_{~~ mn}\hat N^{mn}=0\cr
 &N^{mn} \l^\g D_\g S_{mn,pq}\hat N^{pq}=0~~~~N^{mn}\hat\l^{\hat\g}\hat D_{\hat\g} S_{mn,pq}\hat N^{pq}=0
}}
Since we assumed that the superfields 
$\Omega_{mn,p}, \Omega_{m,pq}, C_{mn}^{~~~~\hat\b}, C^\a_{~~pq}$ and $S_{mn,pq}$
correspond to the linearized curvatures of the connections, we can derive new equations 
needed for the iterative procedure outlined in the introduction. 
By contracting equations \eom\ with respect to the bosonic derivative and antisymmetrizing 
the bosonic indices, one obtains
\eqn\neweom{\eqalign{
& D_\a \Omega_{mn,\hat\b}=\pa_{[m}\g_{n]\a\g}E^\g_{~~\hat\b}~~~~~~~~
\hat D_{\hat\b}\Omega_{\a, pq}=\pa_{[p}\g_{q]\hat\b\hat\g}E_\a^{~~\hat\g}\cr
& D_\a \Omega_{mn,p}=\pa_{[m}\g_{n]\a\g}E^\g_{~~p}~~~~~~~~~\hat D_{\hat\b}\Omega_{m,pq}=-\pa_{[p}\g_{q]\hat\b\hat\g}E_m^{~~\hat\g}\cr
& D_\a C_{mn}^{~~~~\hat\b}=\pa_{[m}\g_{n]\a\g}P^{\g\hat\b}~~~~~~~~~\hat D_{\hat\b}C^\a_{~~pq}=\pa_{[p}\g_{q]\hat\b\hat\g}P^{\a\hat\g}\cr
& D_\a S_{mn,pq} = \pa_{[m}\g_{n]\a\g}C^\g_{~~pq}~~~~~~~\hat D_{\hat\b} S_{mn,pq} =- \pa_{[p}\g_{q]\hat\b\hat\g}C_{mn}^{~~~~\hat\g}
}}
(we define $a_{[m}b_{n]}=a_m b_n-a_n b_m$).  
The identification of the superfields $\Omega_{mn,p}$, $\Omega_{m,pq}$, $C_{mn}^{~~~~\hat\b}$, $C^\a_{~~pq}$ and 
$S_{mn,pq}$ with the linearized curvatures is automatically derived in the formalism \grassi, and 
equations \neweom\ are the usual Bianchi identities.

In order to show that the above equations imply the supergravity
equations of motion we proceed as follows.
We first consider the third line of \neweom\ and 
the $({1\over 2}, {3\over 2}, {3\over 2})$ line of \eom, that we recall for the reader convenience
\eqn\neweomP{\eqalign{
&
D_\a P^{\b\hat\g} -{1\over 4} (\g^{mn})_\a^{~~\b}C_{mn}^{~~~~\hat\g}=0 ~~~~~~~~~~~~ \hat D_{\hat\a} P^{\b\hat\g} - {1\over 4} (\g^{pq})_{\hat\a}^{~~\hat\g}C^{\b}_{~~pq} =0\cr
& D_\a C_{mn}^{~~~~\hat\b}=\pa_{[m}\g_{n]\a\g}P^{\g\hat\b} 
~~~~~~~~~~~~~~~~~~~
\hat D_{\hat\b}C^\a_{~~pq}=\pa_{[p}\g_{q]\hat\b\hat\g}P^{\a\hat\g}\cr
}}  
Acting with $\g^{m}_{\a\sigma} \p_{m}$ on $P^{\sigma \hat \b}$ and 
using the commutation relations of the $D$'s, one gets
\eqn\newI{
\g^{m}_{\a\sigma} \p_{m} P^{\sigma \hat \b} = 
(D_{\a} D_{\sigma} + D_{\sigma} D_{\a}) P^{\sigma \hat \b} 
= {1\over 4}  (\g^{mn})_\a^{~~\sigma} D_{\sigma} C_{mn}^{~~~~\hat\b}\,,
} 
$$
= - {1\over 2}  (\g^{mn})_\a^{~~\sigma} \g_{m \sigma\g} \pa_{n} P^{\g\hat\b} = 
 {9 \over 2}  \g^{m}_{\a\sigma} \p_{m} P^{\sigma \hat \b}
$$
Here we also used the first equation of \neweomP\ and 
$D_{\a} P^{\a\hat\b} =0$ (which follows from \neweomP). In the 
second line we used the first equation in the second line on \neweomP\ and  
the identity $(\g^{mn}\g_{m})_{\a\b} = -9 \g^{n}_{\a\b}$. By performing the 
same manipulations on the hatted quantities we derive the equations
\eqn\newII{
\g^{m}_{\a\sigma} \p_{m} P^{\sigma \hat \b} = 0\,, ~~~~~~~~~~
\g^{m}_{\hat \a \hat\sigma} \p_{m} P^{\a \hat\sigma} = 0\,. 
}
Decomposing $P^{\a\hat \b}$ in terms of Dirac matrices, it is 
straightforward to show that \newII\ implies the equations of motion 
for the RR fields. 

Acting again with $\g^{\a\g}_{n} D_{\a}$ on \newII\ and using equations 
\neweomP\ one gets 
\eqn\newIII{
0 = \g_{n}^{\a\g} \g^{m}_{\a\b} \p_{m} D_{\g} P^{\b\hat\b} = 
(\g_{n}\g^{m})^{\g}_{\b} (\g^{pq})_{\g}^{\b} \p_{m} C_{pq}^{~~~\hat\b} = 
\p^{m} C_{mn}^{~~~\hat\b}\,,
}
and analogously for $C^{\a}_{~~pq}$. These equations are 
the Maxwell equations for the curvature of the gravitinos. They are not 
enough to describe the dynamic of gravitinos and we have to invoke new 
equations coming from the second line of \neweom\ and the 
$({1\over 2}, {3\over 2}, 1)$ line of \neweom. 

Applying $\g^{m}_{\a\sigma} \p_{m}$ on $E^{\sigma}_{~p}$ and 
with  $\g^{p}_{\hat\a\hat\sigma} \p_{p}$ on $E^{~\hat \sigma}_{m}$, the 
same algebraic manipulations yield
\eqn\newIV{
\g^{m}_{\a\sigma} \p_{m} E^{\sigma}_{~p} = 0\,, ~~~~~~~~~~
\g^{p}_{\hat \a \hat\sigma} \p_{p} E_{m}^{~\hat\sigma} = 0\,. 
}
which are the Dirac equations for the gravitinos. These 
equations are gauge invariant under the gauge transformations 
discussed in the next section since the gauge parameters have 
to satisfy a field equation. In addition, 
as above, we find the equations
\eqn\newV{
\p^{m} \Omega_{mn,p} =0\,, ~~~~~~~~~~
\p^{p}  \Omega_{m,pq} =0\,, ~~~~~~~~
}
which are, at the lowest component 
of the superfield $\Omega_{mn,p}$ and $\Omega_{m,pq}$,  
the equations of motion of the graviton, the dilaton and the NS-NS form
\eqn\newVI{
\p^{m} ( \p_{[m} g_{n]p} + \p_{[m} b_{n]p} + \eta_{p[n} \p_{m]}\phi ) = 0\,, 
}
$$
\p^{p} ( \p_{[p} g_{|m|q]} + \p_{[p} b_{|m|q]} + \eta_{nm[q} \p_{p]}\phi) = 0\,. 
$$

Pursuing this line of reasoning, one can derive similar
equations for $E^{\b}_{~~\hat \a}, E_{\a}^{~~\hat \b}, \Omega_{mn,\hat \g}$ 
and $\Omega^{\a}_{~~pq}$, which guarantee that the 
fields are either pure gauge or auxiliary fields. Finally, 
by studying the last line of \neweom\ and the line 
$({1\over 2}, {3\over 2}, 2)$ of \eom, one derives new equations for 
$C^{\b}_{~~pq}, C_{mn}^{~~~~\hat\b}$ and $S_{mn,pq}$, which do not give further information since they are
implied by the previous ones. 


\newsec{Gauge Transformations and Gauge Fixing}

In order to solve the equations of motion \eom\ and \neweom\ it is convenient to choose 
a suitable gauge. Indeed, for supersymmetric theories, the large amount of 
auxiliary fields can be reduced by choosing the Wess-Zumino gauge. 
We first discuss the general structure of the gauge transformations 
\gaugegen, 
we then provide a gauge fixing and we finally check that this gauge can be reached. In the present framework, the gauge parameters 
$\L^{(0,1)}$, $\L^{(1,0)}$, $\tau_z ^{(0,0)}$ and 
$\tau ^{(0,0)}_{\bar z}$ 
satisfy equations \consI\ and \consII\ 
and they are defined up to the gauge transformation \constrans. 
This additional gauge invariance is fixed by a further 
gauge fixing. 

The general structure of the gauge parameters $\L^{(0,1)}$, $\L^{(1,0)}$, $\tau_z ^{(0,0)}$ and 
$\tau ^{(0,0)}_{\bar z}$ is given by
\eqn\gaugeparI{\eqalign{
\L^{(1,0)} = \l^\a \Theta_\a  \quad\quad\quad 
\L^{(0,1)} = \hat \Theta_{\hat\a} \hat\l^{\hat\a}\,,
}}
and 
\eqn\gaugeparII{\eqalign{
& \tau_z^{(0,0)} = \pa_z\theta^\a \Xi_\a + \Pi^m_z \Sigma_m + d_{z\a} \Phi^\a + {1\over 2} N^{mn}_z \Psi_{mn}\cr
& \tau_{\bar z}^{(0,0)} = \hat  \Xi_{\hat\a}\pa_{\bar z}\hat\theta^{\hat\a} + \hat \Sigma_p \hat\Pi^p_{\bar z} + 
\hat \Phi^{\hat\a}\hat d_{\bar z\hat\a} + {1\over 2} \hat \Psi_{pq}\hat N^{pq}_{\bar z}\,.
}}
where $ \Theta_\a, \dots, \hat \Psi_{mn}$ are superfields in the variables $x^m,\t^\a$, and $\hat \t^{\hat \a}$. 
In terms of these superfields, eq. \consI\ gives
\eqn\consIII{(\g^{mnpqr})^{\a\b}D_\b \Theta_\a=0~~~~~~~~~~
(\g^{mnpqr})^{\hat\a\hat\b}\hat D_{\hat\b} \hat \Theta_{\hat\a}=0\,,
}
while eq. \consII\ gives 
\eqn\consIV{\eqalign{
& \Theta_\a + \Xi_\a =0 ~~~~~~~~~~~~~~~~~~~~~~~~~~~~~~ \hat\Theta_{\hat\a}- \hat\Xi_{\hat\a}=0\cr
& D_\a \Theta_\b - D_\b \Xi_\a +\g^m_{\a\b}\Sigma_m=0 ~~~~~~~~~\hat D_{\hat\a}\hat\Theta_{\hat\b} + \hat D_{\hat\b}\hat\Xi_{\hat\a} + \g^p_{\hat\a\hat\b}\hat\Sigma_p=0\cr
& D_\a \Sigma_m +\pa_m \Theta_\a -\g_{m\a\b}\Phi^\b=0 ~~~~~~~ \hat D_{\hat\a}\hat\Sigma_p +\pa_p \hat\Theta_{\hat\a}+ \g_{p\hat\a\hat\b}\hat\Phi^{\hat\b}=0\cr
& D_\a \Phi^\b -{1\over 4} (\g^{mn})_\a^{~~\b}\Psi_{mn}=0~~~~~~~~~~\hat D_{\hat\a}\hat\Phi^{\hat\b} + {1\over 4} \left(\g^{pq}\right)_{\hat\a}^{~~\hat\b}\hat\Psi_{pq}\cr
& N^{mn}\l^\g D_\g \Psi_{mn}=0 ~~~~~~~~~~~~~~~~~~~~~~ \hat\l^{\hat\g}\hat D_{\hat\g}\hat\Psi_{pq}\hat N^{pq}=0\,.
}}
These equations look like the superspace field equations for SYM theory (cf. sec. 1), however 
the superfields $\Theta_\a, \Sigma_m, \Phi^\a$ and $\Psi_{mn}$ and the corresponding hatted quantities depend 
on $x^m, \t^\a$ and $\hat\t^{\hat\a}$. Therefore, the eqs. \consIV\ are not sufficient to determine completely the 
components of those superfields. The free independent components are indeed the gauge parameters.
We also notice that the last pair of equations is trivial when the previous equations and 
the pure spinor conditions are imposed. 
Finally, because of the similarity with SYM case, it is quite natural to impose the condition that $\Psi_{mn}$ and $\hat\Psi_{pq}$ are the linearized curvatures of $\Sigma_m$ and $\hat\Sigma_p$. Again, this assumption is automatic in \grassi. 
The gauge transformations of the superfields in ${\cal V}^{(0,0)}_{z \bar z}$ are given by
\eqn\gaugetrans{\eqalign{
&^{\left(\demi,\demi\right)}~~~~~
\d A_{\a\hat\b}=D_\a \hat\Xi_{\hat\b}+\hat D_{\hat\b}\Theta_\a\cr
&^{\left(\demi,1\right)} ~~~~~~
\d A_{\a p}=\pa_p \Theta_\a +D_\a \hat\Sigma_p;~~~~~~~~~
\d A_{m \hat\b}=\pa_m \hat\Theta_{\hat \b} + \hat D_{\hat\b} \Sigma_m\cr
&^{\left(1,1\right)}~~~~~~
\d A_{mp}=\pa_m \hat\Sigma_p -\pa_p \Sigma_m\cr 
&^{\left({3\over 2},\demi\right)}~~~~~
\d E^\a_{~~\hat\b}= - \hat D_{\hat\b} \Phi^\a; ~~~~~~~~~~~~~~~~~
\d E_\a^{~~\hat\b}=D_\a \hat\Phi^{\hat\b}\cr
&^{\left({3\over 2},1\right)}~~~~~
\d E^\a_{~~p}= -\pa_p \Phi^\a; ~~~~~~~~~~~~~~~~~~~
\d E_m^{~~\hat\b}= \pa_m \hat\Phi^{\hat\b}\cr
&^{\left(\demi,2\right)}~~~~~
\d \Omega_{\a,pq}=D_\a \hat\Psi_{pq};~~~~~~~~~~~~~~~~~~
\d \Omega_{mn,\hat\b}=\hat D_{\hat\b} \Psi_{mn}\cr 
&^{\left(1,2\right)}~~~~~~
\d \Omega_{m,pq}=\pa_m \hat\Psi_{pq};  ~~~~~~~~~~~~~~~~~
\d \Omega_{mn,p}=-\pa_p \Psi_{mn}\cr
&^{\left({3\over 2},{3\over 2}\right)}~~~~~ 
\d P^{\a\hat\b}=0\cr
&^{\left({3\over2},2\right)}~~~~~ 
\d C^{\a}_{~~ pq}=0;~~~~~~~~~~~~~~~~~~~~~~~~~~
\d C_{mn}^{~~~~\hat\b}=0\cr 
&^{\left(2,2\right)}~~~~~~\d S_{mn,pq}=0\,.
}}
From these equations, we easily see that the superfields $P^{\a \hat \b}$, $C^\a_{~~pq}$, $C_{mn}^{~~~~\hat\b}$ and 
$S_{mn,pq}$ are indeed gauge invariant, as expected, being linearized 
field strengths. At zero order in $\t$ and $\hat\t$ eq. \gaugetrans\ gives the gauge transformations of 
supergravity fields. For example, the 
first components of $\hat\Sigma_p = \zeta_p + \xi_p + {\cal O}(\t, \hat\t)$ and 
$\Sigma_m = \zeta_m - \xi_m + {\cal O}(\t, \hat\t)$ are to  be identified with the parameters of 
diffeomorphisms $\delta g_{mp} = \p_m \xi_p + \p_p \xi_m$ and with the gauge transformations 
of the NS-NS form $\delta b_{mp} = \p_m \zeta_p - \p_p \zeta_m$. 
So, the zero-order terms of the gauge parameter superfields $\Theta_\a$, $\Sigma_m$, $\Phi^\a$ 
and of the corresponding hatted quantities are
\eqn\gaugezero{\eqalign{
&\Theta_\a={\cal O}(\t, \hat\t);~~~~~~~~~~~~~~~~~~~~~~\hat\Theta_{\hat\b}={\cal O}(\t, \hat\t)\cr
&\Sigma_m = \zeta_m - \xi_m + {\cal O}(\t, \hat\t);~~~~~~~
\hat\Sigma_p = \zeta_p + \xi_p + {\cal O}(\t, \hat\t)\cr
&\Phi^\a=\varphi^\a+ {\cal O}(\t, \hat\t); ~~~~~~~~~~~~~~~\hat\Phi^{\hat\b}=\hat\varphi^{\hat\b}+ {\cal O}(\t, \hat\t) 
}}
Furthermore, the large amount of gauge parameters allows us to choose the gauge
\eqn\gaugefix{\eqalign{&\t^\a A_{\a\hat\b}=0~~~~~~~~A_{\a\hat\b}~\hat\t^{\hat\b}=0\cr
&\t^\a A_{\a p}=0~~~~~~~A_{m\hat\b}~\hat\t^{\hat\b}=0\cr
& \t^\a E_\a^{~~\hat\b}=0~~~~~~~E^\a_{~~\hat\b}~\hat\t^{\hat\b}=0\cr
&\t^\a~ \Omega_{\a,pq}=0~~~~~\Omega_{mn,\hat\b}~\hat\t^{\hat\b}=0\,.}
}
Indeed, we have at our disposal the parameters $\Theta_\a, \Sigma_m, \Phi^\a$ and $\Psi_{mn}$ and 
the corresponding hatted quantities to impose the gauge \gaugefix. Before showing that the gauge can be reached 
we have to notice that the transformations \gaugetrans~ and the 
equations \consIII\ are invariant under the residual gauge transformations \constrans
\eqn\resid{\eqalign{
&\d \Theta_\a=D_\a \Omega~~~~~~~~~~\d \hat\Xi_{\hat\b}=\hat D_{\hat\b}\Omega\cr
&\d \Sigma_m = - \pa_m \Omega~~~~~~~~\d \hat\Sigma_p= - \pa_p \Omega\cr
&\d \Phi^\a=0~~~~~~~~~~~~~~~\d \hat\Phi^{\hat\b}=0\cr
&\d  \Psi_{mn}=0~~~~~~~~~~~~~\d \hat\Psi_{pq}=0\,,
}}
depending on the scalar superfield $\Upsilon^{(0,0)}=\hat\Upsilon^{(0,0)}\equiv\Omega$.
This requires an additional gauge fixing
\eqn\gaugefixII{
\theta^\a \Theta_\a + \hat\theta^{\hat\b}\hat\Xi_{\hat\b}=0\,.
}

To show that the gauge choice \gaugefix\ can be reached by the gauge transformations 
\gaugetrans, we have to solve, for instance, the equations
\eqn\reacI{
\t^\a (A_{\a\hat\b} + \delta A_{\a\hat\b}) = 0\,, ~~~~~~~~
(A_{\a\hat\b} + \delta A_{\a\hat\b}) \hat\t^{\hat\b} = 0\,,
}
and analogously for all other gauge conditions \gaugefix. By using the properties 
of the superderivative, gauge fixing \gaugefixII, consistency conditions \consIV, and by defining the operators
\eqn\defD{
{\bf D}\equiv \t^\a D_\a=\t^\a{{\pa}\over{\pa\t^\a}}\,,~~~~~~~~{\bf \hat D}\equiv \hat\t^{\hat\b} \hat D_{\hat\b}=\hat\t^{\hat\b}{{\pa}\over{\pa\hat\t^{\hat\b}}}\,,
}
we get the following recursive equations 
\eqn\reacII{\eqalign{
&(1 + {\bf D} + {\bf \hat D}) \Theta_{\a} = - A_{\a \hat \b} \hat\t^{\hat \b}  
- (\g^m \t)_\a \Sigma_m~~~~~
(1 + {\bf D} + {\bf \hat D}) \hat\Theta_{\hat \b} = - \t^\a A_{\a \hat \b} - (\g^p\hat\t)_{\hat\b} \hat\Sigma_p\cr
&({\bf D}+{\bf \hat D})\Sigma_m = A_{m\hat\b}\hat\t^{\hat\b} + (\g_m\t)_\b\Phi^\b~~~~~~~~~~~~~
({\bf D}+{\bf \hat D})\hat\Sigma_p= -\t^\a A_{\a p} - (\g_p\hat\t)_{\hat\g}\hat\Phi^{\hat\g}\cr
&({\bf D}+{\bf \hat D})\Phi^\a = E^\a_{~~\hat\b}\hat\t^{\hat\b} - {1\over 4}(\g^{mn}\t)^\a \Psi_{mn}~~~~~~~
({\bf D}+{\bf \hat D})\hat\Phi^{\hat\b} =- \t^\a E_\a^{~~\hat\b} + {1\over 4}(\g^{pq}\hat\t)^\b \hat\Psi_{pq}\cr
&({\bf D}+{\bf \hat D})\Psi_{mn}=\Omega_{mn,\hat\b}\hat\t^{\hat\b} - (\g_{[m}\t)_\g\p_{n]}\Phi^\g~~~~({\bf D}+{\bf \hat D})\hat\Psi_{pq}=-\t^\a\Omega_{\a,pq}+(\g_{[p}\hat\t)_{\hat\g}\p_{q]}\hat\Phi^{\hat\g}\,.
}}
The operator $({\bf D}+{\bf \hat D})$ acts on homogeneous polynomials 
in $\t^\a$ and $\hat\t^{\hat \a}$ by multiplication by the degree of homogeneity 
and  it does not change its degree. Therefore, the relations \reacII\ are recursive in powers of $\t$ and 
$\hat\t$. 
They can be solved algebraically given $A_{\a \hat \b}, \dots, \Omega_{\a,pq}$ order by order in $\t$ and $\hat\t$ and this proves that the gauge 
can indeed be imposed.
Of course, to reconstruct the gauge-parameter superfields by means of the recursive equations \reacII, we also need lowest order data for them. These are the zero order supergravity gauge parameters \gaugezero.  
 To obtain the last couple of equations we used the additional condition that $\Psi_{mn}$ and $\hat\Psi_{pq}$ are the linearized curvatures of $\Sigma_m$ and $\hat\Sigma_p$.


\newsec{Iterative Procedure and Superfield Reconstruction}

The next step is the derivation of the recursion equations for supergravity superfields. Acting 
with $D_\a$ and $\hat D_{\hat \b}$ on the gauge fixing conditions \gaugefix, and using the 
definition \defD, it is straightforward to derive the recursion relations from eq. \eom
\eqn\recursionDI{\eqalign{
&(1+{\bf D})A_{\a\hat\b}=(\g^m \t)_\a A_{m\hat\b}~~~~~~~~~~(1+{\bf \hat D})A_{\a\hat\b}=(\g^p \hat\t)_{\hat\b}A_{\a p}\cr
&{\bf D}A_{m\hat\b}=(\g_m \t)_\g E^\g_{~~\hat\b}~~~~~~~~~~~~~~~~~~{\bf \hat D}A_{\a p}=(\g_p\hat\t)_{\hat\g}E_\a^{~~\hat\g}\cr
&{\bf D} E^\a_{~~\hat\b}=-{1\over 4} (\g^{mn}\t)^\a \Omega_{mn,\hat\b}~~~~~~~~~{\bf \hat D} E_\a^{~~\hat\b}=- {1\over 4}(\g^{pq}\hat\t)^{\hat\b}\Omega_{\a, pq}\cr
&{\bf D} \Omega_{mn,\hat\b}=-(\g_{[m}\t)_\g \p_{n]} E^\g_{~~\hat\b}~~~~~~~~~{\bf \hat D} \Omega_{\a, pq}=-(\g_{[p}\hat\t)_{\hat\g} \p_{q]} E_\a^{~~\hat\g}
}}
\vskip 12pt

\eqn\recursionDII{\eqalign{
&(1+{\bf D})A_{\a p}= (\g^m \t)_\a A_{mp}~~~~~~~~~~(1+{\bf \hat D})A_{m\hat\b}=- (\g^p \hat\t)_{\hat\b}A_{mp}\cr
&{\bf D}A_{mp}=(\g_m \t)_\b E^\b_{~~p}~~~~~~~~~~~~~~~~~~{\bf \hat D}A_{mp}=-(\g_p\hat\t)_{\hat\b}E_m^{~~\hat\b}\cr
&{\bf D}E^{\a}_{~~p}=-{1\over 4}(\g^{mn}\t)^\a\Omega_{mn,p}~~~~~~~~~~{\bf \hat D}E_m^{~~\hat\b}={1\over 4}(\g^{pq}\hat\t)^{\hat\b}\Omega_{m,pq}\cr
&{\bf D} \Omega_{mn,p}=-(\g_{[m}\t)_\g \p_{n]}E^\g_{~~p}~~~~~~~~~{\bf \hat D} \Omega_{m,pq}=(\g_{[p}\hat\t)_{\hat\g} \p_{q]}E_m^{~~\hat\g}
}}
\vskip 12pt 

\eqn\recursionDIII{\eqalign{
&(1+{\bf D})E_\a^{~~\hat\b}=(\g^m \t)_\a E_m^{~~\hat\b}~~~~~~~~~~(1+{\bf \hat D})E^{\a}_{~~\hat\b}=(\g^p \hat\t)_{\hat\b}E^\a_{~~p}\cr
&{\bf D}E_m^{~~\hat\b}=(\g_m \t)_\g P^{\g\hat\b}~~~~~~~~~~~~~~~~~~~{\bf \hat D}E^\a_{~~p}=(\g_p\hat\t)_{\hat\g}P^{\a\hat\g}\cr
&{\bf D}P^{\a\hat\b}= -{1\over 4} (\g^{mn}\t)^\a C_{mn}^{~~~~\hat\b}~~~~~~~~~~{\bf \hat D}P^{\a\hat\b}= -{1\over 4} (\g^{pq}\hat\t)^{\hat\b}C^\a_{~~pq}\cr
&{\bf D} C_{mn}^{~~~~\hat\b}=-(\g_{[m}\t)_\g \p_{n]}P^{\g\hat\b}~~~~~~~~~~{\bf \hat D} C^\a_{~~pq}=-(\g_{[p}\hat\t)_{\hat\g} \p_{q]}P^{\a\hat\g}
}}
\vskip 12pt
\eqn\recursionDIV{\eqalign{&(1+{\bf D})\Omega_{\a, pq}=(\g^m \t)_\a\Omega_{m,pq}~~~~~~~(1+{\bf \hat D})\Omega_{mn,\hat\b}=-(\g^p\hat\t)_{\hat\b}\Omega_{mn,p}\cr
&{\bf D}\Omega_{m,pq}=(\g_m\t)_\b C^{\b}_{~~pq}~~~~~~~~~~~~~~~{\bf \hat D}\Omega_{mn,p}=-(\g_p\hat\t)_{\hat\b}C_{mn}^{~~~~\hat\b}\cr
&{\bf D}C^{\a}_{~~pq}=-{1\over 4}(\g^{mn}\t)^\a S_{mn,pq}~~~~~~~~{\bf \hat D}C_{mn}^{~~~~\hat\b}={1\over 4}(\g^{pq}\hat\t)^{\hat\b}S_{mn,pq}\cr
&{\bf D}S_{mn,pq}=-(\g_{[m}\t)_\g \p_{n]}C^\g_{~~pq},~~~~~~{\bf \hat D}S_{mn,pq}=(\g_{[p}\hat\t)_{\hat\g}\p_{q]}C_{mn}^{~~~~\hat\g}
}}
A given superfield appears in two groups of equations in order that both its $\t$ and $\hat\t$ components are fixed. 
Inside each group there is an iterative structure (see \har\ and \OoguriPS) which allows us to solve 
those equations recursively given the initial conditions and there is a hierarchical structure among 
the different groups of equations which allows us to solve them subsequently.  
To provide the initial data, we identify the lowest-components of 
the matrix superfield ${\bf A}$ in \vertmatrix\ with supergravity fields 
 \eqn\vertmatrixbc{ 
{\bf A } = 
 \left[ \matrix{ 0 & 0 & 0 & 0 \cr
0 & g_{mp} + b_{mp} + \eta_{mp} \phi 
& \psi_m^{~~\hat\b} & 
\omega_{m, pq} \cr
0 & \psi^\a_{~~p} & f^{\a\hat\b} & 
 {c}^\a_{~~pq}\cr
0 & \omega_{mn,p} & 
 c_{mn}^{~~~~\hat\b} & s_{mn,pq}}\right] + {\cal O}(\t, \hat\t)\,,
 }
where the linearized gravitational connection and curvatures are given by 
\eqn\curva{\eqalign{
&\omega_{m, pq} = (\p_p g_{mq} - \p_q g_{mp}) + (\p_p b_{mq} - \p_q b_{mp}) + (\eta_{mq} \p_p  -  \eta_{mp} \p_q) \phi\,, \cr
&c_{mn}^{~~~~\hat \b} = (\p_m \psi_{n}^{~~\hat \b} - \p_n \psi_{m}^{~~\hat \b})\,,  \cr
&s_{mn,pq} = (\p_m \omega_{n,pq} - \p_n \omega_{m, pq}) \,,
}}
and, analogously, for $\omega_{mn,p}$ and $c^\a_{~~pq}$. 

In the following we give the component-expansion for the physical superfields $A_{mp}$, $E_{m}^{~~\hat\b}$, $E^{\a}_{~~p}$ and $P^{\a\hat\b}$, up to second order in both $\t$ and $\hat\t$. The corresponding curvatures can be easily computed from the defining equations \curva. The component-expansion of the auxiliary superfields is given in appendix B. 
$$\eqalign{
A_{mp} &= (g+b+\eta\phi)_{mp}+ (\g_m\t)_\b\psi^\b_{~~p}-(\g_p\hat\t)_{\hat\b}\psi_m^{~~\hat\b}+(\g_m\t)_\b(\g_p\hat\t)_{\hat\g}f^{\b\hat\g} \cr
&-{1\over 8}(\g_m\t)_\b (\g^{nr}\t)^\b\omega_{nr,p} -{1\over 8}(\g_p\hat\t)_{\hat\b}(\g^{qr}\hat\t)^{\hat\b}\omega_{m,qr}\cr
&+{1\over 8}(\g_m\t)_\b(\g^{nr}\t)^\b(\g_p\hat\t)_{\hat\g}c_{nr}^{~~~\hat\g}-{1\over 8}(\g_m\t)_\g(\g_p\hat\t)_{\hat\b}(\g^{qr}\hat\t)^{\hat\b}c^\g_{~~qr}\cr
&+{1\over 64}(\g_m\t)_\b(\g^{nr}\t)^\b (\g_p\hat\t)_{\hat\g}(\g^{qs}\hat\t)^{\hat\g}s_{nr,qs}+\dots \cr
& }$$
\eqn\compo{\eqalign{
E_{m}^{~~\hat\b} &= \psi_m^{~~\hat\b} + (\g_m\t)_\g f^{\g\hat\b}+{1\over 4}(\g^{pq}\hat\t)^{\hat\b}\omega_{m,pq}-{1\over 4}(\g_m\t)_\g (\g^{pq}\hat\t)^{\hat\b} c^\g_{~~pq}\cr
&-{1\over 8}(\g_m\t)_\g (\g^{nr}\t)^\g c_{nr}^{~~~\hat\b} + {1\over 4}(\g^{pq}\hat\t)^{\hat\b}(\g_p\hat\t)_{\hat\g}\pa_q\psi_m^{~~\hat\g}\cr
&-{1\over 32}(\g_m\t)_\g(\g^{nr}\t)^\g(\g^{pq}\hat\t)^{\hat\b}s_{nr,pq}+{1\over 4}(\g_m\t)_\g(\g^{pq}\hat\t)^{\hat\b}(\g_p\hat\t)_{\hat\g}\pa_q f^{\g\hat\g}\cr
&-{1\over 32}(\g_m\t)_\g (\g^{nr}\t)^\g (\g^{pq}\hat\t)^{\hat\b}(\g_p\hat\t)_{\hat\g} \pa_q c_{nr}^{~~~\hat\g}+\dots\cr
& 
}}
$$
\eqalign{
E^{\a}_{~~p} &= \psi^\a_{~~p} - {1\over 4}(\g^{mn}\t)^\a\omega_{mn,p} + (\g_p\hat\t)_{\hat\g}f^{\a\hat\g}+ {1\over 4}(\g^{mn}\t)^\a(\g_p\hat\t)_{\hat\b} c_{mn}^{~~~\hat\b}\cr
&+{1\over 4}(\g^{mn}\t)^\a (\g_m\t)_\g\pa_n\psi^\g_{~~p} - {1\over 8}(\g_p\hat\t)_{\hat\g}(\g^{qr}\hat\t)^{\hat\g} c^\a_{~~qr}\cr
&+{1\over 4}(\g^{mn}\t)^\a(\g_m\t)_\g (\g_p\hat\t)_{\hat\b}\pa_n f^{\g\hat\b}+{1\over 32}(\g^{mn}\t)^\a(\g_p\hat\t)_{\hat\g}(\g^{qr}\hat\t)^{\hat\g}s_{mn,qr}\cr
&+{1\over 32}(\g^{mn}\t)^\a (\g_m\t)_\g(\g_p\hat\t)_{\hat\b}(\g^{qr}\hat\t)^{\hat\b} \pa_n c^\g_{~~qr} +\dots\cr
& 
}$$
$$\eqalign{
P^{\a\hat \b} &= f^{\a\hat\b} - {1\over 4}(\g^{mn}\t)^\a c_{mn}^{~~~~\hat\b}-{1\over 4}(\g^{pq}\hat\t)^{\hat\b}c^\a_{~~pq}- {1\over 16}(\g^{mn}\t)^\a (\g^{pq}\hat\t)^{\hat\b}s_{mn,pq}\cr
&+{1\over 4}(\g^{mn}\t)^\a (\g_m\t)_\g \pa_n f^{\g\hat\b} +{1\over 4}(\g^{pq}\hat\t)^{\hat\b}(\g_p\hat\t)_{\hat\g} \pa_q f^{\a\hat\g}\cr
&+{1\over 16}(\g^{mn}\t)^\a(\g_m\t)_\g (\g^{pq}\hat\t)^{\hat\b}\pa_n c^\g_{~~pq}-{1\over 16}(\g^{mn}\t)^\a(\g^{pq}\hat\t)^{\hat\b}(\g_p\hat\t)_{\hat\g}\pa_q c_{mn}^{~~~\hat\g}\cr
& +{1\over 16}(\g^{mn}\t)^\a (\g_m\t)_\g(\g^{pq}\hat\t)^{\hat\b}(\g_p\hat\t)_{\hat\g}\p_n \p_q f^{\g\hat\g}+\dots
}$$
It is easy to verify that this expansion satisfies the gauge conditions \gaugefix\  and that all 
auxiliary fields have been eliminated and reexpressed in terms of derivatives of physical supergravity fields.

The next step is to insert the expansion \compo\ into the definition of the vertex operator \vert\ and 
recombine the worldsheet one-forms ${\bf X}_z$ and ${\bf X}_{\bar z}$ in order to 
get a more manageable expression. However, it makes sense to provide such expression for an interesting example in sec. 7.  

We have to notice that the vertex operator ${\cal V}^{(1,1)}$ contains only the superfield 
$A_{\a \hat \b}$ which encodes all the needed information regarding the supergravity fields, which however
appear at higher orders in $\t$'s and $\hat\t$'s. This is sufficient for amplitudes computations, even 
though the measure factor on zero modes in the correlation functions has to 
soak up plenty of $\t$'s and $\hat\t$'s (\BerkovitsFE, \BerkovitsPH,  and \GrassiNZ). 


\newsec{Gauge Fixing for Massive States}

In the previous sections, we explored the gauge fixing 
for the massless sector of open and closed string theory. However, 
the spectrum of string theory contains infinitely many massive states 
defined, in the closed string case, by the equations 
\eqn\maA{
 \Big[Q_L, {\cal V}^{(1,1)}_{n}\Big] = 0 \,,
  ~~~~~
 \Big[Q_R, {\cal V}^{(1,1)}_{n}\Big] = 0\,,
  ~~~~~
\Big[L_{0,L} + L_{0,R} -n, {\cal V}^{(1,1)}_{n}\Big]= 0 \,,
}
where $L_{0,L} = \oint dz~ z~ T_{zz}$ and $L_{0,R} = \oint d\bar z~ \bar z ~\bar T_{\bar z\bar z}$. 
The index $n$ denotes the mass of the state. Even if these equations can be solved by 
expanding the vertex operators ${\cal V}^{(1,1)}_{n}$ in terms of the building-blocks 
$\p \t^{\a}$, $\bar\p \hat\t^{\hat\a}$, $\Pi^{m}$, $\bar \Pi^{m}$,..., it is convenient to fix 
a gauge as in the massless case 
and then solve the equations by an iterative construction 
as shown in the previous section. However, since we cannot explore the complete set of vertices and provide a gauge fixing for each of them, we propose a definition of gauge fixing based on new anticommuting and 
nilpotent charges to be imposed 
on the physical states. This resembles the Siegel gauge (where 
the corresponding charges are $b_{L,0}  = \oint dz \,z \,b_{zz}$ and 
$b_{L,0}  = \oint d\bar z \,\bar z \, \hat b_{\bar z\bar z}$ where 
$b_{zz}$ and $\hat b_{\bar z\bar z}$ are the left- and right-moving 
antighosts) used in string field theory to eliminate all auxiliary fields and 
to define the propagator for the string field. 

We introduce the following charges ``dual'' to the BRST operators
\eqn\maB{
{\cal K}_{L} = \oint dz \, \t^{\a} w_{\a}\,, ~~~~~
{\cal K}_{R} = \oint d\bar z \, \hat\t^{\hat \a} \hat w_{\hat \a}\,.
} 
They are nilpotent and anti-commute. They are not supersymmetry invariant 
 as can be directly seen by the presence of $\t^{\a}$ and $\hat \t^{\hat \a}$. This in fact 
 implies that we are choosing a non symmetric gauge which can be viewed as a generalization 
 of the Wess-Zumino gauge condition in 10 dimensions. It eliminates the lowest non physical 
 component of the superfields and it fixes the auxiliary fields -- appearing at higher order 
 in the superspace expansion -- in terms of the physical fields and their derivatives. In addition, 
 ${\cal K}_{L/R}$ are not invariant under the gauge transformations (A.4) , but 
 their gauge variations are BRST invariant because of the pure spinor conditions
 \eqn\maBB{
 \{Q_{L}, \Delta_{L} {\cal K}_{L} \} = 0 \,, ~~~~~~
 \{Q_{R}, \Delta_{R} {\cal K}_{R} \} = 0 \,, }
 Moreover,  ${\cal K}_{L/R}$ have  
 the following commutation relations with the BRST operators
 \eqn\maC{
 \{Q_{L}, {\cal K}_{L}\} =  {\cal D} + J_{L}\,, ~~~~~
 \{Q_{R}, {\cal K}_{L} \} = 0\,,
  }
$$
\{Q_{R}, {\cal K}_{R}\} =  \hat{\cal D} + J_{R}\,, ~~~~~
 \{Q_{L}, {\cal K}_{R} \} = 0\,,
 $$
where 
\eqn\maD{
{\cal D} = \oint dz : \t^{\a} d_{\a} :\,, ~~~~~~ J_{L} = \oint dz : \l^{\a} w_{\a} :
 }
$$
\hat{\cal D} = \oint d\bar z :\hat \t^{\hat \a} \hat d_{\hat \a} :\,, ~~~~~~ 
J_{R} = \oint d\bar z : \l^{\hat \a} \hat w_{\hat \a} :
$$
Acting on superfields $F(x,\t, \hat \t)$, 
we have that  $\{ {\cal D}, F\} = {\bf D} F$ and 
$\{ \hat{\cal D}, F\} = {\bf \hat D} F$. 
The ordering of fields in the operators ${\cal D}, \hat{\cal D}, J_{L}$ and 
$J_{R}$ is needed to define the corresponding currents. The operators are 
gauge invariant under (A.4) because of \maBB. The main difference 
with respect to Siegel gauge fixing in string field theory is that in that case  
$b_{zz}$ and $\hat b_{\hat z \hat z}$ are 
holomorphic and antiholomorphic anticommuting currents of 
spin 2. 

In the case of the open superstring, denoting by $Q$ and by ${\cal K}$ the BRST and gauge fixing operators, the gauge condition on the massless vertex operator ${\cal V}^{(1)} = \l^{\a} A_{\a}$ is given by 
\eqn\maE{
\{{\cal K}, {\cal V}^{(1)} \} = \oint dw \, \Big(\t^{\a} w_{\a} \Big)(w) \, \Big(\l^{\a} A_{\a}(x,\t)\Big)(z) = \t^{\a} A_{\a} = 0\,.
}
We notice that the field $\t^{\a}$ in ${\cal K}$ is 
harmless for massless vertices, but it will give a nontrivial contribution in the massive case. In the latter 
case one has to add a compensating non-gauge invariant contribution on the 
r.h.s. of \maE\ in order to compensate the fact that ${\cal K}$ is not gauge invariant 
under (A.4).

Applying $Q$ on the left hand side of \maE,
applying ${\cal K}$ on the equation 
$\{Q, {\cal V}^{(1)}\} = \l \g^{m} \l A_{m}(x,\t)=0$  
and using the commutation 
relations \maC, we obtain
\eqn\maEE{
({\bf D} + 1) {\cal V}^{(1)} = \l \g^{m} \t A_{m}\,. 
}
Eliminating the ghost $\l^{\a}$, we end up with 
equation \III\ for the superfields $A_{\a}$ and $A_{m}$. This 
procedure can be clearly generalized to massive states. First, we 
discuss the closed string case, then we show an example for the 
first massive state for open superstrings and, finally, we show that 
the zero momentum cohomology satisfies the same equations generalized 
to zero modes. 

For closed strings, we reproduce the gauge fixing \gaugefix\ by
the following conditions 
\eqn\maF{
\{{\cal K}_{L}, {\cal V}^{(1,1)} \} = 0\,, ~~~~~~  \{{\cal K}_{R}, {\cal V}^{(1,1)} \} = 0
}
and, for the gauge parameters $\Lambda^{(1,0)}$ and $\Lambda^{(0,1)}$ in eq. 
\gaugeparI, by the gauge condition
\eqn\maG{
\{ {\cal K}_{L}, \L^{(1,0)}\} + \{ {\cal K}_{R}, \L^{(0,1)} \} = 0\,.
}
which coincides with \gaugefixII. Applying the BRST 
charge on the left hand sides of \maF, acting 
with ${\cal K}_{L}$ and ${\cal K}_{R}$ on equations \deC, 
and finally using the commutation relations \maC, we derive the conditions 
for the iterative equations given in the previous section. 

Let us show that the gauge fixing \maE\ also fixes the gauge transformations in a suitable way for the first massive state of the open superstring ${\cal V}^{(1)}_{1}$, leading to a 
recursive procedure to compute the vertex operator in term of the initial data, a 
multiplet of on-shell fields containing a massive spin 2 field \GS. 

A general decomposition of ${\cal V}^{(1)}_{1}$ in terms of fundamental building-blocks is given by 
\eqn\maH{
{\cal V}^{(1)}_{1} = \p\l^{\a}A_{\a} + \l^{\a} \p\t^{\b}  B_{\a\b} 
+ \l^{\a} :\Pi^{m} C_{\a m}: + \l^{\a} :d_{\b} D^{\b}_{~~\a}: 
}
$$
+:\l^{\a} N^{mn}: E_{\a mn} + :\l^{\a} w_{\b} \l^{\b}: F_{\a} 
$$
and its gauge transformation is generated by 
\eqn\maI{
\delta {\cal V}^{(1)}_{1} =\Big[ Q, \Omega^{(0)}_{1} \Big]\,,
}
with
$$
\Omega^{(0)}_{1} =  \p\t^{\b}  \Omega_{\b} 
+ :\Pi^{m} \Omega_{m}: + :d_{\b} \Omega^{\b}: 
+:N^{mn}: \Omega_{mn} + :w_{\b} \l^{\b}: \Omega \,. 
$$
The decompositions are based on the requirement that the 
vertex operator should be invariant under the gauge transformation $\Delta$ 
given in (A.4). A further gauge transformation of $\Omega^{(0)}_{1}$ 
would be a variation of a negative ghost number field. The only one is the antighost 
$w_{\a}$, but there is no gauge invariant operator only with $w_{\a}$ without $\l^{\a}$. Notice 
that we have to add a (BRST-invariant) 
compensating term of the form $w \g^{mnpq} \l$ in order to reabsorb the non-invariance of 
${\cal K}$. 

Imposing \maE, we get 
\eqn\maJ{
A_{\a} + \t^{\b} B_{\b\a} = 0\,, ~~~~~~~ \t^{\a} C_{\a m} = 0\,,~~~~~~~ \t^{\b} D_{\b}^{~\a} = 0 \,,
}
$$
\t^{\b} E_{\b mn} + {1\over 1440}\left[(\g_{mn})^{\a}_{~\g} D^{\g}_{~~\a}-(\g_{mn})^{\a}_{~\g}\t^\g F_\a\right] = 0\,, ~~~~~~
{1\over 2}(\g^{mn}\t)^\b E_{\b mn}-D^{\a}_{~~\a}+2\t^\a F_\a = 0\,. 
$$
This gauge fixing can be reached by adjusting the parameters 
$\Omega_{\a}, \Omega_{m}, \Omega^{\a}, \Omega_{mn}$ and $\Omega$. Using 
 equations \maJ\ and applying the operator ${\bf D}$, we obtain 
the iterative relations to compute the vertex. 
The gauge fixing \maJ\ fixes only the supergauge part of the 
gauge transformation. This gauge 
does not fix the physical gauge transformation of the massive spin 2 system \BerkovitsQX. 

Finally, we show that the measure for zero modes satisfies the gauge fixing proposed 
above. In fact, by restricting the attention to zero momentum cohomology, 
we supersede ${\cal K}$ with the differential 
\eqn\maZ{
{\cal K}_{0} = \t^{\a}_{0} {\partial \over \partial \l^{\a}_{0}} 
} 
which acting on ${\cal V}^{(3)} = ( \l_{0}\g^{m}\t_{0}) 
(\l_{0}\g^{n}\t_{0})( \l_{0}\g^{p}\t_{0})( \t_{0} \g_{mnp}\t_{0}) $, 
yields 
\eqn\maZZ{
{\cal K}_{0} {\cal V}^{(3)} = 0\,. 
}
Similarly, for the closed superstring, the 
ghost number $(3,3)$ cohomology representative 
satisfies the corresponding gauge fixing. 

Even if the gauge fixing is not manifestly supersymmetric, 
the supersymmetry of the target space theory is still a symmetry. 
As usual, in the Wess-Zumino gauge, a supersymmetry 
transformation must be accompanied by a gauge transformation to 
bring the vertex to the original gauge. This means that 
\eqn\susA{
\delta_{\e} [{\cal K}, {\cal V}] + [{\cal K}, \delta {\cal V}] =0 
}
where $\delta {\cal V} = [Q, \Omega_{\e}]$, $\delta_{\e} {\cal V} = 
[ \e^{\a} Q_{\a}, {\cal V}]$ and $Q_{\a} = \oint dz \, q_{\a}$ (the 
supersymmetry generator $q_{\a}$ is given in (A.8)).  
As an example, we 
show that $\Omega_{\e}$ can be indeed found for 
the massless sector of the open superstring and the extension 
is similar for the other cases. Equation \susA\ reduces 
to 
\eqn\susB{
\e^{\a} A_{\a} + \t^{\a} \e^{\b} Q_{\b} A_{\a} + \t^{\a} D_{\a} \Omega_{\e} =0\,, 
}
which yields
\eqn\susC{
{\bf D} \Omega_{\e} = 0\,.
}
Again, this equation can be solved iteratively in powers of $\t$'s and 
it follows that $\Omega = \Omega_{0}(x)$. \susC\ can be checked explicitly 
on the solutions \compo. 


\newsec{Non-Constant RR Field-Strength}

In \noncom\ sigma models for superstrings 
in the presence of constant RR field strengths have been studied.
It has been verified that 
non-(anti)commutative superspaces \NCsuperspace\ 
naturally appear in the presence of that background. 
A series of applications, 
from topological strings to deformed supersymmetric instanton analysis, has then been considered 
\NACappl.  In \noncom, it has also been conjectured that from non-constant 
RR field strengths one can derive new equal-time commutation 
relations between coordinates $x^{m}$ and $\t^{\a}$ living on the 
boundaries such as 
\eqn\nnon{
\{\t^{\a}, \t^{\b} \} = \g^{\a\b}_{m} x^{m}\,, 
 }
 generalizing the construction of Lie-algebraic 
 non-commutative geometries to supermanifolds \pvn~ (for a different example of a Lie-algebraic geometry in superspace see \lie). 
 
 The vertex operator 
 for non-constant RR fields strengths is the basic ingredient of this kind of analysis. 
 Applying our method, we 
 compute the vertex for linearly $x$-dependent RR field strengths

\eqn\defapprox{
P^{\a\hat\b}=f^{\a\hat\b}+{\cal C}_m^{~~\a\hat\b}x^m} 
where ${\cal C}_m^{~~\a\hat\b}$ is constant. $P^{\a\hat\b}$ must satisfy eqs. \newII, which become $\g^{m}_{\a\b} {\cal C}_{m}^{\b\hat\g} = \g^{m}_{\hat\a\hat\b} {\cal C}^{\g \hat\a} =0$ for the specific ansatz \defapprox \foot{
Equations \newII\ can be rewritten in terms of forms by decomposing $P^{\a\hat\b}$ according to 
Dirac equations. For example, for type IIB we have the 1-form $P_{m}$, the 3-form $P_{[mnp]}$ and 
the 5-form $F_{[mnpqr]}$. Solving the Bianchi identities we get $P_{m} = \p_{m} A$, $ P_{[mnp]} = \p_{[m} A_{np]}$,... 
and the field equations are $\p^{m} P_{m} = \p^{2} A = 0$, $\p^{m}  P_{[mnp]} = \p^{m} \p_{[m} A_{np]}$,... 
These can be solved in terms of quadratic polynomials 
$A(x) = (10 \, a_{(mn)} - a_{r}^{~r} \eta_{mn}) x^{m} x^{n}$, $A_{[mn]} = (10 \, a_{[mn],(rs)} - 
a_{[mn,t]}^{~~t} \eta_{rs})  x^{r} x^{s}$,... where $a_{(mn)}$, $a_{[mn], (rs)}$,... are 
constant background fields. 
}. 
 
We remind the reader that, in the constant field strength case, ${\cal V}_{z\bar z}^{(0,0)}=q_\a f^{\a\hat\b} \hat q_{\hat\b}$,
where $q_\a$ and $\hat q_{\hat\b}$ are the supersymmetry currents given in (A.8).
So it is easy to see that equation \deE\ is verified with
${\cal V}^{(1,1)}=\chi_\a f^{\a\hat\b}\hat\chi_{\hat\b}$, which is
clearly BRST invariant (see (A.9) and (A.10)). 
Since in the $\t$ and $\hat\t$ expansions of the auxiliary and physical superfields $A_{\a\hat\b}$,...,$P^{\a\hat\b}$ (see eqs. \compo\ and appendix B) 
the number of bosonic derivatives acting on physical zero-order components grows with growing order in $\t$ and $\hat\t$, it is clear that the ansatz \defapprox\ will correspond to only a few non-zero terms in the expansion.
Actually, the highest-order contributions are $\t^4\hat\t^2$ and $\t^2\hat\t^4$ terms. Here 
we give the explicit expressions for $A_{\a\hat\b}$ 
\eqn\nonconstRRII{
\eqalign{ 
A_{\a\hat \b}& = {1\over 9}(\g^m\t)_\a (\g_m\t)_\b (\g^p\hat\t)_{\hat\b}(\g_p\hat\t)_{\hat\g}(f^{\b\hat\g}+{\cal C}_n^{~~\b\hat\g}x^n)\cr
&+{1\over 180}(\g^m\t)_\a(\g_m\t)_\g(\g^p\hat\t)_{\hat\b}(\g_p\hat\t)_{\hat\g}(\g^{qr}\hat\t)^{\hat\g}(\g_q\hat\t)_{\hat\d}{\cal C}_r^{~~\g\hat\d}\cr
&+{1\over 180}(\g^m\t)_\a(\g_m\t)_\d(\g^{nr}\t)^\d(\g_n\t)_\g(\g_p\hat\t)_{\hat\b}(\g^p\hat\t)_{\hat\g}{\cal C}_r^{~~\g\hat\g} \,, 
}}
The remaining superfields are given in app. C.  
To obtain the vertices ${\cal V}^{(1,1)}$ and ${\cal V}^{(0,0)}_{z\bar z}$ for the linearly $x$-dependent RR field strength we have to insert 
\nonconstRRII\ and the superfields given in app. C back into \verB\ and \vert.

For the unintegrated vertex operator we find 
\eqn\Unonconst{\eqalign{{\cal V}^{(1,1)}&=\chi_\a f^{\a\hat\b}\hat\chi_{\hat\b}\cr
&+\chi_\a\left[\left(x^m \d_\g^\a \d_{\hat\g}^{\hat\b}+{1\over 20}(\g^{qm}\hat\t)^{\hat\b}(\g_q \hat\t)_{\hat\g} \d_\g^\a + {1\over 20}(\g^{nm}\t)^\a (\g_n\t)_\g \d_{\hat\g}^{\hat\b}\right){\cal C}_m^{~~\g\hat\g}\right]\hat\chi_{\hat\b}}}
while for the integrated vertex operator ${\cal V}^{(0,0)}_{z\bar z}$ we obtain
\eqn\Vnonconst{\eqalign{{\cal V}^{(0,0)}_{z\bar z}&=q_\a f^{\a\hat\b} q_{\hat\b}\cr
&+q_\a\left[x^s \d_\g^\a \d_{\hat\g}^{\hat\b} + {1\over 4}(\g^{rs}\t)^\a (\g_r\t)_\g \d_{\hat\g}^{\hat\b} + {1\over 4} (\g^{ps}\hat\t)^{\hat\b}(\g_p\hat\t)_{\hat\g}\d_\g^\a\right]{\cal C}_s^{~~\g\hat\g}\hat q_{\hat\b}\cr
&+ \left[-{1\over 6}(\p x^m + {1\over 10}\t\g^m\p\t)(\t\g_m\g^{rs}\t)-N^{rs}\right](\g_r\t)_\a {\cal C}_s^{~~\a\hat\b}\hat q_{\hat\b}\cr
&+q_\a {\cal C}_s^{~~\a\hat\b}(\g_r\hat\t)_{\hat\b}\left[-{1\over 6}(\bar\p x^p + {1\over 10}\hat\t \g^p\bar\p\hat\t)(\hat\t\g_p\g^{rs}\hat\t)-\hat N^{rs}\right]
}}

The complicated structure of ${\cal V}^{(0,0)}_{z\bar z}$ prevents 
from a simple analysis of superspace deformations as in \noncom, 
and this will be discussed in a separate publication \lref\prepa{P.A. Grassi and 
L. Tamassia, { in preparation.}} \prepa. 


\newsec{Vertex Operators with RR Gauge Potentials}

In presence of D-branes, one can ask which states couple to them and 
which vertex operators describe such interaction. As it was discussed in  \asymI\ 
in the framework of RNS formalism, one has to construct the 
vertex operators for RR fields in the asymmetric picture.\foot{We 
thank M. Bianchi for suggesting the present analysis.} In addition, 
a propagating closed string ({\it i.e.} with non vanishing momentum) 
emitted from a disk or a D-brane, has to be off-shell. Therefore, 
one needs to break the BRST invariance by allowing 
a non vanishing commutator with $Q_{L, 0} + Q_{R,0}$ where 
$Q_{L/R, 0}$ are the picture conserving parts of BRST charges in the 
RNS formalism. In particular in \asymII\ the authors construct a 
solution of $[ Q_{L, 1} + Q_{R,1}, W]=0$, where $W$ is the vertex operator 
in the asymmetric picture. 

In the present section, we construct analogous vertices for 
closed superstrings which do not satisfy the classical supergravity 
equations of motion, but modified superfield constraints. 
They allow a description of the 
RR gauge potentials, in contradistinction to the on-shell formalism case, 
where only the field strengths $P^{\a\hat \b}$ 
appear. First of all, there are some important differences. The 
two BRST charges $Q_{L}$ and $Q_{R}$ contain a single 
term and therefore the decomposition used in \asymII\ is 
not viable. In addition, there are no different 
 pictures (in the usual sense) for a given vertex since there are no superghosts 
 associated to local worldsheet supersymmetry.  
 There is, however, the possibility of constructing two operators 
 which resemble the picture lowering and raising operator, as 
 suggested by Berkovits \pricom, but the implications 
 of this new idea have not been explored yet. 
 
Nevertheless we can construct an off-shell formalism 
by considering the 
following combination of vertices with different ghost numbers: 
\eqn\vI{
{\cal V}^{(2)} = {\cal V}^{(2,0)} + {\cal V}^{(1,1)} + {\cal V}^{(0,2)}\,,
}
where the notation ${\cal V}^{(a,b)}$ stands for vertex operators 
with the left ghost number $a$ and with the right ghost number $b$. 
The ghost number of the l.h.s. is just the sum of the ghost numbers. 
Notice that if we insert a single ${\cal V}^{(2)}$ vertex in an amplitude where 
all the other vertices have definite ghost number, only the 
central part ${\cal V}^{(1,1)} $ of the vertex ${\cal V}^{(2)}$ does contribute. In principle, 
we have to add all the possible terms $\sum_{i = - \infty}^{\infty}  {\cal V}^{(1+i,1-i)}$. 
However, only the antighosts $w_{z \a}$ and $\hat w_{\bar z\hat \a }$, which have to be gauge invariant under (A.4), carry a 
negative ghost number and higher conformal spin. This is not the 
case in the RNS formalism where the presence of a worldsheet index is compensated 
by the worldsheet ghosts that carry conformal spin $-1$. 

The general form of ${\cal V}^{(2)}$ is given by 
\eqn\vII{
{\cal V}^{(2)} = 
\l^{\a} \l^{\b} G_{(\a\b)} + 
\l^{\a} \hat\l^{\hat\b} A_{\a\hat\b} +
\hat\l^{\hat\a} \hat\l^{\hat\b}\hat G_{(\hat\a\hat\b)} \,. 
}
where $G_{(\a\b)}$, $A_{\a\hat\b}$ and $\hat G_{(\hat\a\hat\b)}$ are 
superfields of the variables $x^{m}, \t^{\a}$ and $\hat\t^{\hat \a}$. The 
superfields $G_{(\a\b)}$ and $\hat G_{(\hat\a\hat\b)}$ are 
symmetric in the spinorial indices and 
they are defined up to the algebraic gauge transformations 
\eqn\vIII{
\delta G_{(\a\b)} = \g^{m}_{\a\b} \Sigma_{m}\,, ~~~~~~~~
\delta \hat 
G_{(\hat\a\hat\b)} = \g^{m}_{\hat\a\hat\b} \hat\Sigma_{m}\,, ~~~~~
}
which leave only the 5-form parts unfixed. 

Instead of imposing the equations \deC\ separately 
for the left and the right sector, we then impose the modified condition
\eqn\vIV{
[ Q_{L} + Q_{R}, {\cal V}^{(2)} ] = 0\,,
} 
which leads to the equations
\eqn\vV{\eqalign{
& D_{\a} A_{\b\hat\b} + D_{\b} A_{\a\hat\b} - \g^{m}_{\a\b} A_{m \hat\beta} =- \hat D_{\hat\b} G_{(\a\b)}\,, 
\cr
& \hat D_{\hat\a} A_{\a\hat\b} + \hat D_{\hat\b} A_{\a\hat\a} - \g^{m}_{\hat\a\hat\b} 
A_{\a m} =- D_{\a} \hat G_{(\hat\a\hat\b)}\,, 
\cr 
& D_{\a} G_{(\b\g)} + D_{\b} G_{(\g\a)} + D_{\g} G_{(\a\b)} =\g^{m}_{(\a\b} G_{\g) m} \,, 
\cr
& 
\hat D_{\hat\a} \hat G_{(\hat\b\hat\g)} + 
\hat D_{\hat\b} \hat G_{(\hat\g\hat\a)} + 
\hat D_{\hat\g} \hat G_{(\hat\a\hat\b)} =\g^{m}_{(\hat\a\hat\b}\hat G_{\hat\g)m} \,, 
}}
The first two equations are consistent generalizations of the first 
two equations in \eom. The other two equations emerge from the ghost number three 
part of \vIV. They are equations of motion for the auxiliary fields $G_{(\a\b)}$ and $\hat G_{(\hat \a \hat \b)}$. 
It has been shown in the context of spinorial cohomology 
\spincoho\ 
that the third and fourth equations have non-trivial solutions. Therefore, 
the first and second equations are deformations of the usual superfield constraints. In the limit $G, \hat G \rightarrow 0$ one recovers the usual on-shell supergravity. 
It is interesting to note that the way supergravity equations and/or 
on-shell superspace constraints are relaxed in order to go off-shell follows very closely the case of N=1 SYM theory \SYM\ 
presented in \spincoho. In fact, in those papers the authors considered an 
extension of d=10 N=1 SYM theory by relaxing the 
superspace constraint $F_{(\a\b)} \equiv D_{(\a} A_{\b)} - \g^{m}_{\a\b} A_{m} = 0$ 
(cf. the introduction for the notation) by introducing a self-dual 5-form
$F_{(\a\b)} = J^{[5]}_{(\a\b)}$. In the present case, the relaxation 
of the on-shell constraints is again due to two 5-forms $G_{(\a\b)}$ and $\hat G_{(\hat \a \hat \b)}$. 
The 5-form 
$J^{[5]}_{(\a\b)}$ satisfies the constraint $D_{(\a} J^{[5]}_{\b\g)} - \g^{m}_{(\a\b} J_{\g) m} =0$ 
due to the Bianchi identities. The solution of these identities by means of the 
spinorial cohomology shows that the 5-form is not completely constrained, but 
it allows extensions of the superspace equation of motions. 
In the same way, the conditions for $G_{(\a\b)}$ and $\hat G_{(\hat \a \hat \b)}$ constrain 
them, but allow a certain useful modifications of the supergravity equations 
of motion. 

Eqs. \vV\ are invariant under the gauge transformations 
\eqn\vVI{
\delta {\cal V}^{(2)} = [ Q_{L} + Q_{R}, \Omega^{(1)}]\,, ~~~~~
\Omega^{(1)} = \l^{\a} \Theta_{\a} + \hat\l^{\hat \a} \hat\Theta_{\hat \a} \,.
}
which imply
\eqn\vVII{
\delta A_{\a\hat \b} = D_{\a} \hat\Theta_{\hat\b} + \hat D_{\hat \b} \Theta_{\a}\,, ~~~~~
}
$$
\delta G_{(\a\b)} = \g^{m}_{\a\b} \Sigma_{m} + D_{(\a} \Theta_{\b)}\,, ~~~~~
\delta \hat G_{(\hat\a\hat\b)} = \g^{m}_{\hat\a\hat\b} 
\hat\Sigma_{m} + \hat D_{(\hat\a} \hat\Theta_{\hat\b)}\,. 
$$
In the last equations we have summed \vIII\ to 
\vVI. The gauge parameters $\Theta_{\a}$ and
$\hat\Theta_{\hat \a}$ are not forced to satisfy consistency conditions 
as in \consI\ and \consII. Nevertheless, 
eqs. \vV\ are gauge invariant because of the 
second and the third equations of  \vVII.

Following the previous section, we can require a suitable gauge fixing 
to reduce the auxiliary fields by imposing
\eqn\vVIIA{
\{ {\cal K}_{L} + {\cal K}_{R}, {\cal V}^{(2)} \} =0 .
}
This gauge condition yields
\eqn\vVIIB{
2 \t^{\b} G_{(\a \b)} + \hat \t^{\b} A_{\a\hat \b} =0\,, ~~~~~
\t^{\a} A_{\a\hat \b} +  2 \hat \t^{\hat \a}\hat G_{(\hat \a \hat \b)} =0\,,
}
which partially fix \vVII. Notice that the lowest component 
of $A_{\a\hat \b}$ is chosen to be kept unfixed and \vVIIB\ can be 
regarded as a condition on $G$ and $\hat G$. In this gauge the lowest 
component of $A_{\a\hat\b}$ is not gauged away and 
for example we can identify the $\t^{2}$ and $\hat\t^{2}$ component 
of $A_{\a\hat\b}$ as the RR potentials
\eqn\RRpo{
A_{\a\hat\b} = \dots + a_{\a}^{~~\hat \g} (\g_{m}\hat\t)_{\hat\g} (\g^{m}\hat\t)_{\hat\b} + 
 \hat a^{\g}_{~~\hat \b} (\g_{m}\t)_{\g} (\g^{m}\t)_{\a} + \dots 
 }
where $a_{\a}^{~~\hat\g}$ and $\hat a^{\a}_{~~\hat \g}$ are the RR potentials. 
Those potentials can also be found in the lowest components of 
the superfields $E_{\a}^{~~\hat \g}$ and $E^{\a}_{~~\hat \g}$. 

As a consistency condition we observe that 
\eqn\vVIIB{
\{ Q_{L} + Q_{R}, {\cal K}_{L} + {\cal K}_{R} \} = 
({\cal D} + \hat {\cal D}) + ( J_{L} + J_{R})\,, 
}
and by using \vVI\ and \vVIIA\ we obtain a consistency condition 
for ${\cal V}^{(2)}$. It can be checked that the vertex ${\cal V}^{(2)}$ 
satisfies this new constraint and this condition does 
not fix the physical gauge transformations.

The natural question is how to extend the analysis of the previous 
sections in the present case. We have to rederive the complete set of 
descent equations in order to provide all the equations of motion. 
The generalized vertex operators are given by the collection 
\eqn\vVIII{
{\cal V}^{(1)}_{\bar z} = {\cal V}^{(2,-1)}_{\bar z} + {\cal V}^{(1,0)}_{\bar z} + {\cal V}^{(0,1)}_{\bar z}  \,,
}
$$
{\cal V}^{(1)}_{z} = {\cal V}^{(1,0)}_{z} + {\cal V}^{(0,1)}_{z} + {\cal V}^{(-1,2)}_{z}  \,,
$$
$$
{\cal V}^{(0)}_{z \bar z} = {\cal V}^{(1,-1)}_{z\bar z} + {\cal V}^{(0,0)}_{z\bar z} + 
{\cal V}^{(-1,1)}_{z \bar z}  \,. 
$$
It is easy to show that all equations \eom\ are deformed by the 
presence of new fields. By setting those fields to zero 
we recover the on-shell supergravity equations. 


\newsec{Antifields and the Kinetic Terms for Closed String Field Theory}

In the present section, we derive the set of antifields 
for the massless sector of closed string theory. We discuss the couplings 
of the fields to the antifields for a closed 
string field theory action and, finally, we propose a kinetic term 
which leads to the correct equations of motion taking 
into account the presence of selfdual forms. This section 
is structured as follows. We first recall some basic facts 
about the field theory of open superstrings, mainly 
focusing on the relation between fields and antifields, their coupling
and the kinetic terms yielding the linearized equations 
of motion.   
Following the analogy with closed bosonic string field theory, we then
construct the antifields and the kinetic terms for closed 
superstrings. Since in the present 
paper we never dealt with non-linear extensions of supergravity 
equations, we do not discuss generalizations of 
Witten's string field $\star$-product for open superstring. 
Similarly, it is outside of the scope of the present paper 
to construct a full-fledged closed string field theory and we limit 
ourselves to a specific sector of the theory. 


\subsec{Open Superstring (Antifields and the Kinetic Term)} 

One of the most important ingredients in the construction 
of an off-shell extension of superstring theory is the 
Batalin-Vilkovisky formalism. Associated to each field of the theory $\varphi_{s}$ 
(the index $s$ denotes a collection of fields with the same ghost 
number $G(s)$),
there is an antifield $\varphi^{*}_{s}$, 
whose BRST variation 
corresponds to the equations of motion of $\varphi^{*}$. The 
set of antifields for open superstring theory at the massless 
level has been discussed in \berko\ 
(see also \chest). The 
generalization at the massive level is straightforward, 
and the ghost-for-ghosts for higher massive spin fields has 
to be taken into account. 
 
Following the notation of the previous sections, 
we introduce the string field $\Phi^{(1)}_{o}$ 
which has the general decomposition (at the massless level) 
\eqn\sfA{
\Phi^{(1)}_{o} = 
C + \l^{\a} A_{\a} + \l^{\a} \l^{\b} A^{*}_{(\a\b)} + \l^{\a} \l^{\b}
 \l^{\g} C^{*}_{(\a\b\g)}\,.
} 
The truncation at order three in the ghost fields is justified by the absence 
of any cohomology at ghost number greater than three\foot{There is a
simple way to establish the absence of cohomology at ghost number higher than 
three by using supergeometrical arguments discussed in \Belo\ and references therein.} and the 
only cohomology at ghost number three is the zero momentum 
cohomology constructed in terms of 
\eqn\sfB{
C^{*}_{\a\b\g} = C^{*} 
(\g^{m}\t)_{\a} (\g^{n}\t)_{\b} (\g^{r}\t)_{\g} (\t \g_{mnr} \t) 
}
where $C^{*}$ is constant. The expansion of $\Phi^{(1)}_{o}$ 
into powers of $\l^{\a}$ accounts for different target space 
fields: the ghost superfield  $C$, the spinorial part of the 
connection $A_{\a}$, the antifields $A^{*}_{(\a\b)}$ and the 
antifield $C^{*}_{(\a\b\g)}$ of the ghost fields. The interpretation 
as target space fields with different ghost and antifield numbers 
is discussed in \siegel\ and as rigid symmetries of the 
target space in 
\lref\BerkovitsGJ{
N.~Berkovits, M.~T.~Hatsuda and W.~Siegel,
Nucl.\ Phys.\ B {\bf 371}, 434 (1992)
[hep-th/9108021].
} \BerkovitsGJ.
In order to assign a total (worldsheet plus target 
space) ghost number to $\Phi^{(1)}_{0}$, the target space ghost numbers must be  
+1 for the field $C$, $0$ for $A_{\a}$, $-1$ for $A^{*}_{\a\b}$, and 
$-2$ for $C^{*}_{\a\b\g}$. They coincide precisely with the 
usual assignment of ghost and antifield number for theories
with irreducible gauge symmetries. The ghost number zero superfield $A_{\a}$ 
contains the gauge field $a_{m}(x)$ and the gluino $\psi^{\a}(x)$, the 
ghost number $-1$ superfield $A^{*}_{\a\b}$ contains the antifields 
$a^{*,m}$ of $a_{m}(x)$ and $\psi^{*}_{\a}$ of $\psi^{\a}(x)$. 

The main properties are the following: 

{\it i)} The string field
 $\Phi^{(1)}_{o} = \Phi^{+}_{o} + \Phi^{-}_{o}$ is  
decomposed into fields $\Phi^{+}_{o}$ and antifields $\Phi^{-}_{o}$ which have 
the expansions 
\eqn\osfA{
\Phi^{+}_{o} = \sum_{s, G(s) \geq 0} \varphi_{s} \Phi^{+}_{o,s}\,, ~~~~~~
\Phi^{-}_{o} = \sum_{s, G(s) < 0} \varphi^{*}_{s} \Phi^{-}_{o,s}\,. }
where $\Phi^{+}_{o,s}$ and $\Phi^{-}_{o,s}$ form two basis for 
fields and antifields, respectively.  

If the inner product $\langle A, B \rangle = 
\int d\mu^{(-3)} A B$ is defined by using the measure\foot{An expression 
for $d\mu^{(-3)}$ is given in \GrassiNZ.}
\eqn\osfB{
\int d\mu^{(-3)} {\cal V}^{(3)}_{0} =1\,, ~~~~~~
{\cal V}^{(3)}_{0} = (\l_{0} \g^{m} \t_{0})( \l_{0} \g^{n} \t_{0})( \l_{0} \g^{r} \t_{0})( \t_{0} \g_{mnr} \t_{0})\,,
} 
the antifields $\Phi^{-}_{o}$ are dual to the fields $\Phi^{+}_{o}$ according to the 
relation 
\eqn\osfC{
\int d\mu^{(-3)} \Big(\Phi^{+}_{o} \Phi^{-}_{o}\Big) = \sum_{s,s'} \int d\mu^{(-3)} 
(\varphi_{s} \Phi^{+}_{o,s}) (\varphi^{*}_{s'} \Phi^{-}_{o,s'}) = 
\sum_{s} \int d^{10}x \, \varphi_{s} \varphi^{*}_{s}\,.
}
The integration over the Grassman coordinates 
is prescribed by the measure $d\mu^{(-3)}$. 

{\it ii)} The antifields $\Phi^{-}_{o}$ couple in the action
to the BRST variation of  fields. Therefore, the unique choice is 
\eqn\osfD{
S^{*} = \int d\mu^{(-3)} \Big( \Phi^{-}_{o} Q \Phi^{+}_{o}  \Big)
\,.
}
It is easy to check that the antifields $\varphi^{*}_{s}$ correctly couple to the 
variation of the fields $\varphi_{s}$. A term of the form $
S^{*} = \int d\mu^{(-3)} \Big( \Phi^{-}_{o} Q \Phi^{-}_{o}  \Big)$ 
vanishes because of the ghost number of $\Phi^{-}_{s}$. 

{\it iii)} Finally, a kinetic term for the fields can be constructed. 
Since the inner product defined above is non-degenerate, one can check that 
the kinetic term 
\eqn\osfE{
S^{K} = \int d\mu^{(-3)} \Big( \Phi^{+}_{o} Q \Phi^{+}_{o} \Big)\,,
}
yields the correct linearized equations of motion. It has been 
verified that, for massless fields, the above equation leads to the 
correct component action for N=1 SYM in $d=(9,1)$. 
It would be very interesting 
to check if \osfE\ also leads to an action for free massive higher spin fields  
\BerkovitsUA (see also 
\lref\SiegelYZ{
W.~Siegel,
{\it Introduction To String Field Theory,}
Adv.\ Ser.\ Math.\ Phys.\  {\bf 8}, 1 (1988).
hep-th/0107094.
} \SiegelYZ). 

. 


\subsec{Closed Superstrings (Antifields and the Kinetic Term)}

We base our construction on the bosonic closed string 
field theory. We follow the notations and the definitions 
given in \zwie\ and we provide a translation in terms of our formulation. 

In closed string theory  a 
generic string field $\Phi^{(1,1)}$ has ghost number $(1,1)$ and the usual 
form for the kinetic term $\int d\mu^{(-3,-3)} \, \Phi Q \Phi $ given above cannot work. 
In fact the BRST operator $Q_{L}+ Q_{R}$ has ghost number $(1,0) + (0,1)$, 
the measure has ghost number $(-3,-3)$ and 
it is not possible to saturate the ghost number properly. In \zwie, 
it was proven that the kinetic term for bosonic closed 
string theory 
\eqn\csfA{
S_{c} = \langle \Phi^{(1,1)} | c^{-}_{0} (Q_{L}+ Q_{R})| \Phi^{(1,1)} \rangle \,,
}
where $c^{-}_{0} = c_{L,0} - c_{R,0}$ ($c_{L/R,0}$ are the 
zero modes of diffeomorphisms ghosts) leads to the correct equations of motion. 
Moreover, string field theory actions have the property to 
reproduce not only the action for the physical fields (in 
the case of bosonic closed string field theory, eq. \csfA\ 
yields the action for the graviton, for the NS-NS two form and 
for the dilaton), but also the full BV action 
with antifields and gauge transformations. 

As explained before, removing the restriction of the ghost number of the string field $\Phi_{c}$ 
\foot{We add the subscript $c$ to distinguish it from the open case and we remove the 
ghost number.} we can decompose it 
into $\Phi_{c} = \Phi^{+}_{c} + \Phi^{-}_{c}$ 
where $\Phi^{+}_{c} = \sum_{s, G(s) \geq 0} \varphi_{s} \Phi^{+}_{s}$ 
with $G(s)$ denoting the ghost number of field 
$\varphi_{s}$. 
 The other components of $\Phi_{c}$, namely $\Phi_{c}^{-}$, should 
contain the antifields. 

The antifields should be dual with respect to the 
inner product given in \csfA\ $\langle A |c_{0}^{-}| B \rangle$, 
where $A$ and $B$ are two generic vertex operators.\foot{In \zwie, it 
has been proved that the inner product is not degenerate. However, 
the gauge invariance is achieved only if the string field $\Phi_{c}$ satisfies 
the two conditions: $\{b_{0}^{-} , \Phi_{c}\} = \{L^{-}_{0}, \Phi_{c}\} =0$. 
} 
Since the fields, the ghosts and the ghost-for-ghosts 
are contained into $\Phi_{c}^{+}$ with ghost number $0,1$ and 
$2$ respectively, the corresponding antifields $\Phi^{-}_{0}$ 
should have ghost number $3,4$ and $5$ to be dual to 
$\Phi^{+}_{c}$. 
In fact for bosonic closed strings, 
in order to get a non zero result from the zero mode prescription 
\eqn\bzero{
\langle c_{L,0} c_{L,1}  c_{L,-1} c_{R,0} c_{R,1} c_{R,-1} \rangle =1\,,
}
for tree level amplitudes, the total ghost number should be 6.  
(On the sphere, one has to compensate the anomaly 
of the left- and right-moving ghost current anomaly).
However, the naive BPZ conjugation of a string field maps 
$\Phi_{c}$ into 
$\tilde\Phi_{c} = \sum_{s, G(s) < 0} 
\varphi^{*}_{s} \tilde\Phi_{s}$. So, in addition, one has to act 
with the operator $b^{-}_{0}$ (where $b_{L/R,0}$ are the 
zero modes of the antighosts) 
\eqn\csfD{
\Phi^{-}_{c} = \Big\{b^{-}_{0} , \sum_{s, G(s) < 0} 
\varphi^{*}_{s} \tilde\Phi_{s} \Big\} = 
\sum_{s, G(s) < 0} 
\varphi^{*}_{s} \{b_{0}^{-}, \tilde\Phi_{s}\} 
=
\sum_{s, G(s) < 0} 
\varphi^{*}_{s} \Phi^{-}_{s} 
\,,
}
The $b_{0}^{-}$ operation has the 
virtue to  correctly reduce the ghost number of the conjugated 
string field. In particular, $\Phi^{+}_{s}$, defined above, forms a
basis and $\Phi^{-}_{s}$ forms the dual basis 
paired with $\Phi^{+}_{s}$ as 
\eqn\cfsE{
\langle \Phi^{-}_{s} | c_{0}^{-} |\Phi^{+}_{s'} \rangle = \delta_{s,s'}\,.
} 
Removing the restrictions on ghost numbers, one 
can show that there is always an element of the basis 
$\Phi^{-}_{c}$ corresponding to an element of $\Phi^{+}_{c}$. 

In the present framework, although we have 
neither the operator $b_{0}^{-}$ nor the operator $c^{-}_{0}$, 
we can still construct the 
antifields $\sum_{s} \varphi^{*}_{s} \Phi^{-}_{s}$ paired to 
the fields $\sum_{s} \varphi_{s} \Phi^{+}_{s}$. In particular for the 
massless sector $\Phi^{+}_{s}$ has the general expansion 
\eqn\csfB{
 \Phi^{+}_{c}  = \Omega + 
  \l^{\a}\Theta_{\a} + \hat\l^{\hat \a}\hat \Theta_{\hat\a} + 
 \l^{\a} \hat\l^{\hat \b} A_{\a\hat\b}\,.}
where $A_{\a\hat\b}$ contains the fields, $\Theta_{\a}$ and $\hat\Theta_{\hat \a}$ 
contain the target space ghosts and $\Omega$ the ghost-for-ghosts. 
\foot{For massive states, we have also
 to take into account the expansion in the antighost 
 $w_{\a}$ as in \maH.} 
Being the measure for zero modes of $\l^{\a}, \hat\l^{\hat\a}, \t^{\a}$ 
and $\hat\t^{\hat\a}$ given by the following equations 
\eqn\zmA{
\int d{\mu}^{(-3,-3)}_{c} {\cal V}^{(3,3)} = 1\,, ~~~~~~
 }
$$
 {\cal V}^{(3,3)} =  
 (\l_{0}\g^{m}\t_{0} 
\l_{0}\g^{n}\t_{0} \l_{0}\g^{p}\t_{0} \t_{0} \g_{mnp}\t_{0}) 
(\hat\l_{0}\g^{m} \hat\t_{0} 
\hat\l_{0}\g^{n}\hat\t_{0} \hat\l_{0}\g^{p}\hat\t_{0} \hat\t_{0} \g_{mnp}\hat\t_{0}) \,,
$$
it follows that the components of the expression 
\eqn\csfF{
\Phi^{-}_{c} = 
\l^{\a} \l^{\b} \hat\l^{\hat \a} \hat\l^{\hat \b} A^{*}_{(\a\b)(\hat \a \hat\b)} + 
\l^{\a} \l^{\b} \l^{\g} \hat\l^{\hat \a} \hat\l^{\hat \b} 
\Theta^{*}_{(\a\b\g)(\hat \a \hat\b)}
}
$$
+ 
\l^{\a} \l^{\b}  \hat\l^{\hat \a} \hat\l^{\hat \b} \hat\l^{\hat \g} 
\hat\Theta^{*}_{(\a\b)(\hat \a \hat\b \hat\g)} + 
\l^{\a} \l^{\b} \l^{\g} \hat\l^{\hat \a} \hat\l^{\hat \b} \hat\l^{\hat \g}
\Omega^{*}_{(\a\b\g)(\hat \a \hat\b \hat \g)}\,,
$$
are indeed paired by \zmA\ to the different components 
$\Phi^{+}_{c}$ (in the massless sector). 
The superfields 
$A^{*}_{(\a\b)(\hat \a \hat\b)}, 
\Theta^{*}_{(\a\b\g)(\hat \a \hat\b)}, 
\hat\Theta^{*}_{(\a\b)(\hat \a \hat\b \hat\g)}$ and 
$\Omega^{*}_{(\a\b\g)(\hat \a \hat\b \hat \g)}$ are defined 
up to the algebraic gauge transformations 
\eqn\csfG{\eqalign{
&\delta A^{*}_{(\a\b)(\hat \a \hat\b)} 
= 
\g^{m}_{\a\b} A^{*}_{m,(\hat \a \hat\b)} + 
\g^{m}_{\hat\a\hat\b} A^{*}_{\a \b, m}\,, \cr
&\delta \Theta^{*}_{(\a\b\g)(\hat \a \hat\b)} 
=
\g^{m}_{(\a\b} \Theta^{*}_{\g) m,(\hat \a \hat\b)} + 
\g^{m}_{(\hat\a\hat\b} \Theta^{*}_{|\a \b|,\hat \g) m}\,, \cr
&\delta \Theta^{*}_{(\a\b\g)(\hat \a \hat\b \hat\g)} 
=
\g^{m}_{(\a\b} \Theta^{*}_{\g) m,(\hat \a \hat\b \hat\g)} + 
\g^{m}_{(\hat\a\hat\b} \Theta^{*}_{|\a \b \g|,\hat\g) m}\,, 
}}
which reduce the number of independent components in order 
to match the number of fields present in $\Phi^{+}_{c}$. 
A long, but straightforward computation shows that
\eqn\csfH{ 
\int d\mu^{(-3,-3)} 
\Big(\Phi^{-}_{c} \Phi^{+}_{c} \Big) = \int d^{10}x 
\sum_{s} 
\varphi^{*}_{s} \varphi_{s}\,,
} 
where the Grassman integration is performed according to the measure 
given in \zmA. 

{\it ii)} In the next step, we use the above definition to 
construct a term which relates the antifields to the BRST 
sources of the physical fields. Since the gauge transformations 
are generated by the BRST charge $Q_{L}+Q_{R}$, 
the term of the action which couples antifields and BRST variations 
is 
\eqn\csfI{S^{*} =
\int d\mu^{(-3,-3)} 
\Big( \Phi^{-}_{c} (Q_{L} + Q_{R}) \Phi^{+}_{c} \Big) \,.
} 
The BRST operator raises the ghost number of the string field $\Phi^{+}_{c}$ of one unity, therefore 
\eqn\csfJ{
S^{*} =  
\sum_{s} \int d^{10}x \varphi^{*}_{s} \delta \varphi_{s} 
}
where $\delta \varphi_{s}$ is the BRST variation of the 
field $\varphi_{s}$ which depends on the fields $\varphi_{s-1}$.  

{\it iii)} As a last step, we have to construct the kinetic term of the closed 
string field theory. Since we do not have the operators $b_{0}^{-}$ and 
$c^{-}_{0}$, we cannot follow the bosonic string field theory analogy. 
However, in our context we have a new vertex 
operator with ghost number 3 of the form 
\eqn\csfK{
\Phi^{(3)} = 
\l^{\a} \l^{\b} \hat \l^{\hat \a}  {\cal Q}_{(\a\b) \hat \a} + 
\l^{\a}  \hat \l^{\hat \a} \hat \l^{\hat \b} \hat {\cal Q}_{\a \hat \a \hat \b}\,,  
}
which has been neglected so far. 
${\cal Q}_{(\a\b) \hat \a}$ and $\hat{\cal Q}_{\a (\hat \a \hat \b)}$ are defined up to the
gauge transformations $\delta {\cal Q}_{(\a\b) \hat \a} = \g^{m}_{\a\b} \Omega_{m \hat \a}$ and 
$\delta \hat{\cal Q}_{\a (\hat \a \hat \b)} = \g^{m}_{(\hat \a\hat \b)} \Omega_{\a m}$ which allow 
us to take into account the pure spinor constraints. 

We can finally write a kinetic term of the form 
\eqn\csfL{
S^{K} = \int d\mu^{(-3,-3)} 
\Big(
\Phi^{(3)}_{0} (Q_{L} + Q_{R}) \Phi^{+}_{c} 
 + \sum_{n \geq 0} (-1)^{n} \Phi^{(3)}_{n} \Phi^{(3)}_{n+1} \Big) \,,  
 }
where $\Phi^{(3)}_{n}$ is an infinite collection 
of fields of the form \csfK\ needed to escape the no-go theorem 
of the existence of an action for selfdual antisymmetric forms with
a finite number of fields and covariant
 \dualactions. 
 To write action \csfL\ we followed the technique 
 developed in \BerkoHull\ and it has similarity 
 with action proposed in \BerkovitsGJ\ where 
 all the pictures are taken simultaneously into account 
 introducing new commuting fields. The expansion 
 of the string fields in terms of powers of these new fields 
 leads to a target space action with infinite number of fields 
 \BerkoCSFT. 
 
 The action \csfL\ is invariant under the 
gauge transformations 
\eqn\cfsLA{
\delta \Phi^{+}_{c} = (Q_{L} +Q_{R}) \Omega + \Delta\,, ~~~~~
}
$$
\delta \Phi^{(3)}_{2n} = (Q_{L} +Q_{R}) \Gamma\,, ~~~~~
\delta \Phi^{(3)}_{2n +1} = (Q_{L} +Q_{R}) \Delta\,, ~~~\forall~n.
$$
where $\Omega$ has total ghost number 1, whereas $\Delta$ and 
$\Gamma$ have ghost number 2.
Notice that there is 
no coupling between $\Phi^{(3)}_{n}$ and the antifields, because of the ghost number of $\Phi^{(3)}_{n}$. 

The presence of commuting ghosts $\l^{\a}$ and $\hat \l^{\hat \a}$ is usually the 
source of a well-known problem, an infinite number of equivalent copies of the cohomology, 
identified by a new quantum number known as picture. In the context 
of topological theory the Picture Changing Operator (PCO) and its 
inverse have been constructed on the basis of supergeometry and singular forms 
\Belo. In pure spinor string theory, Berkovits has recently suggested that the PCO and 
its inverse can be indeed constructed 
\pricom. This would motivate the introduction of an infinite number of fields $\Phi^{(3)}_{n}$ 
necessary to construct a string field theory 
action for closed superstrings. 

From the action $S^{K}$ one can derive the equations of motion 
\eqn\csfM{
\{Q_{L}+ Q_{R}, \Phi^{+}_{c} \} + \Phi^{(3)}_{1} = 0\,, 
~~~~~~~~\{Q_{L} + Q_{R}\, \Phi^{(3)}_{0} \} = 0 \,, }
$$
\Phi^{(3)}_{0} = \Phi^{(3)}_{2} = \Phi^{(3)}_{4} = \dots \,,
$$
$$
\Phi^{(3)}_{1} = \Phi^{(3)}_{3} = \Phi^{(3)}_{5} = \dots \,.
$$
and for a solution with a finite number of string fields, one gets 
\eqn\csfM{
\{Q_{L}+ Q_{R}, \Phi^{+}_{c} \} = 0\,, 
}
$$
\Phi^{(3)}_{0} = \Phi^{(3)}_{2} = \Phi^{(3)}_{4} = \dots  =0 \,, 
$$
$$
\Phi^{(3)}_{1} = \Phi^{(3)}_{3} = \Phi^{(3)}_{5} = \dots 
=0\,.
$$
which are the equations of motion discussed in the previous 
sections. This also confirms the fact that the fields 
$\Phi^{(3)}_{n}$ are not propagating and they are only 
Lagrange multipliers. For a solution with finite number of fields, the 
equations are invariant under the gauge transformations
with $\Delta =0$ and $(Q_{L}+Q_{R}) \Gamma=0$. This coincides with the 
correct gauge invariance of the theory. 

The action given summing \csfJ\ and \csfL\ has several properties which 
justify its form: {\it i)} it leads to the correct equations of motion in the 
space of solutions with a finite number of  fields, {\it ii)} it has the correct gauge 
transformations and it is explicitly supersymmetric, {\it iii)} the relation between 
fields and antifields is realized, {\it iv)} the new operators $\Phi^{(3)}_{n}$ play only 
an auxiliary role and they do not propagate, {\it v)} the action avoids the no-go theorems about 
selfdual forms, {\it vi)} the action suggests a Chern-Simons-like action in a higher ``dimension''
 (that can be reduced to the form in \csfJ\ by discretizing the new dimension) and can be interpreted as 
a Chern-Simons action for the supermembrane where one has to reabsorb 7 ghosts 
(see  \BerkovitsUC).  

\newsec{Outlook}

We collected, analysed and studied some of the fundamental ingredients for amplitude 
computations in covariant superstring formalism. We showed the power of the 
present framework, where we were able to 
construct a systematic procedure to compute the 
vertex operators for closed covariant superstrings. We also found a way to relax superspace contraints and we proposed a tentative closed string field theory action. However, there are several open questions which were not addressed in this paper: 
{\it a)} the extension of amplitude computations beyond tree level, {\it b)} 
the computation of deformed superspaces associated to non constant RR fields \prepa, 
{\it c)} the analysis of T-duality (it is rather simple to check T-duality for 
vertex operators, but a detailed analysis should be done at the level of sigma 
model \howe, {\it d)} the role played by conformal invariance and worldsheet 
diffeomorphisms, last but not least {\it e)} a full-fledged field theory for 
open and closed strings. 
To answer some of these questions one should probably follow a more 
geometrical approach based on WZW actions 
\lref\GrassiKQ{
P.~A.~Grassi, G.~Policastro and P.~van Nieuwenhuizen,
Nucl.\ Phys.\ B {\bf 676}, 43 (2004)
[hep-th/0307056].
}
\lref\GrassiCZ{
P.~A.~Grassi and P.~van Nieuwenhuizen,
hep-th/0403209.
}
\lref\GuttenbergHT{
S.~Guttenberg, J.~Knapp and M.~Kreuzer,
hep-th/0405007.
}
\refs{\GrassiKQ,\GrassiCZ,\GuttenbergHT}.


\vskip .5cm
\hskip -.7cm  {\bf Acknowledgments}
\vskip .5cm

This work was partly funded by NSF Grant PHY-0098527. The work of L.T. is supported by INFN and MURST. 
P.A.G. thanks the Theory Division  at CERN and IHES (Bures-sur-Yvette) 
for financial support.
We thank N. Berkovits, M. Bianchi, L. Castellani, A. Lerda, P. van Nieuwenhuizen, S. Penati, R. Russo, W. Siegel and A. Zaffaroni for useful discussions and suggestions.


\vskip .5cm 

\appendix{A}{BRST Symmetry, Gauge Invariance and the Sigma Model}

The field content is $x^{m}$ where $m=0,\dots,9$, two Majorana-Weyl spinors $\t^{\a},\hat \t^{\hat \a}$ 
with $\a=\hat \a =1, \dots, 16$ and their conjugate momenta $\p x_{m}$, $p_{\a}$ and $\hat p_{\hat \a}$. The Dirac 
matrices $\g^{m}_{\a\b}$ and $\g^{m}_{\hat\a\hat\b}$ are the 
$16\times 16$ off-diagonal blocks of $Spin(9,1)$ Dirac matrices. 
They are real and symmetric and they satisfy the Fierz identities 
$\g^{m}_{\a(\b} \g_{m \g\d)} =0$. We introduce the commuting Weyl spinors 
$\l^{\a}$ and $\hat\l^{\hat \a}$, which satisfy the pure spinor conditions 
$$\l
 \g^{m} \l =0\,, ~~~~~
  \hat\l \g^{m} \hat\l=0\,,$$ 
  and their conjugate momenta $w_{\a}, \hat w_{\hat\a}$. The solution of the pure 
  spinor constraints can be only achieved by breaking Lorentz invariance, however we do not 
  need to solve them in the present paper.  
It is very important to introduce the supersymmetric invariant composite operators 
\eqn\ccA{
d_{\a} = p_{\a} - {1\over 2} \p x^{m} (\g_{m} \t)_{\a} - {1\over 8} (\g^{m} \t)_{\a} (\t \g_{m} \p\t)
\,, }
$$
\hat d_{\hat\a} = \hat p_{\bar\a} - {1\over 2} \bar\p x^{m} (\g_{m} \hat\t)_{\hat\a} - 
{1\over 8} (\g^{m} \hat\t)_{\hat \a} (\hat\t \g_{m} \bar\p\hat\t) 
\,, 
$$

Following Berkovits, we define the BRST operators 
\eqn\conC{
Q_{L} = \oint dz \l^{\a} d_{\a}\,, ~~~~~~~~ Q_{R} = \oint d\bar z \hat\l^{\hat\a} \hat d_{\hat\a}\,.
}
which satisfy
\eqn\conCA{
Q_{L}^{2} = -\oint dz\, \l\g^{m}\l \Pi_{m}\,, ~~~~~
[Q_{L}, Q_{R}]=0\,, ~~~~
Q_{R}^{2} =- \oint d\bar z\, \hat\l\g^{m}\hat\l \hat\Pi_{m}\,, ~~~~~
}
where $\Pi_z^m = \p x^m + {1\over 2}\t \g^m \p \t$ and 
$\hat\Pi_{\bar z}^m = \bar\p x^m +{1\over 2} \hat\t \g^m \bar\p \hat\t$. 

Due to pure spinor constraints, they 
are nilpotent up to gauge transformations of $w_{\a}, \hat w_{\hat\a}$, 
given by 
\eqn\conD{
\Delta_{L} w_{\a} = \L_{m} (\g^{m} \l)_{\a}\,, ~~~~~~~~
\Delta_{R} \hat w_{\a} = \hat \L_{m} (\g^{m} \hat\l)_{\a}\,.
}
with 
the local parameters $\L_{m}$ and $\hat \L_{m}$ generated by the pure spinor constraints. 
These gauge transformations remove the degrees of freedom 
from the covariant $w_{\a}$ and $\hat w_{\hat \a}$ 
 the independent dof of the pure spinors $\l^{\a}$ and $\hat\l^{\hat \a}$. 
 Gauge invariant operators are 
 \eqn\gi{
 {\cal J}_{L} = : w_{\a} \l^{\a} :\,, ~~~~~~~  {\cal J}_{R} = : \hat w_{\a} \hat\l^{\a} :\,, ~~~~~~~ 
 }
$$
{N}_{L} =  {1\over 2}:w \g^{mn} \l :\,, ~~~~~~~  {N}_{R} =  {1\over 2}:\hat w \g^{mn} \hat\l :\,, ~~~~~~~ 
$$

Following the usual prescription of the BRST 
quantization rules, we can define the quantum action as follows
\eqn\conDA{
S_{0} =  S_{GS} + 
Q_{L} \int d^{2}z w_{\a} \bar\p \t^{\a} + Q_{R} \int d^{2}z \hat w_{\hat\a} \p \hat\t^{\hat\a}\,.
}
where $S_{GS}$ is the Green-Schwarz action in the conformal gauge \GS. 
Even if this looks like the usual BRST procedure, we have to notice that 
the BRST-like operators $Q_{L}$ and $Q_{R}$ are nilpotent up to gauge 
transformations \conD. This compensates the fact that the Green-Schwarz 
action is not invariant under BRST transformations. In addition, we can always add BRST invariant terms to the 
action. However, there is no procedure 
to get \conDA\ from an honest gauge fixing of the Green-Schwarz action 
(a suggestion is given in \tonin). 

By exploiting the different contributions in \conDA, we obtain 
\eqn\conE{
S_{0} = \int d^2 z 
\Big({1\over 2}\p x^m \bar \p x_m + p_\a \bar \p\t^\a +\hat p_{\hat\a}\p \hat\t^{\hat\a}+ w_\a \bar\p \l^\a + \hat w_{\hat\a}\p \hat\l^{\hat\a}  \Big)\,,
 }
which is BRST invariant and invariant under the gauge transformation \conD\ if the spinors 
$\l^{\a}, \hat\l^{\hat\a}$ are pure. The action is also invariant under supersymmetry 
transformations generated by $Q_{\e} = \e^{\a} \oint dz q_{\a} + 
\hat \e^{\hat \a} \, \oint d\bar z \hat q_{\hat \a}$
where the explicit expressions for the supersymmetry currents are
\eqn\conF{
q_{\a} = p_{\a} + {1\over 2} \p x^{m} (\g_{m} \t)_{\a} + {1\over 24}(\t \g^{m} \p\t) (\g_{m} \t)_{\a} 
\,, }
$$
\hat q_{\hat\a} = \hat p_{\bar\a} +{1\over 2} \bar\p x^{m} (\g_{m} \hat\t)_{\hat\a} + 
{1\over 24}(\hat\t \g^{m} \bar\p\hat\t)  (\g_{m} \hat\t)_{\hat \a} 
\,. 
$$
These do not anticommute with the BRST operators $Q_L$ and $Q_R$, since
\eqn\susyQ{[Q_L,q_\a]=\p \chi_\a\,,~~~~~~~~~~[Q_R,\hat q_{\hat\b}]=\bar\p \hat\chi_{\hat\b}}
where $\chi_\a$ and $\hat\chi_{\hat\b}$ are the BRST-invariant quantities 
\eqn\susyQII{\chi_\a\equiv{1\over 3}(\l\g^m\t)(\g_m\t)_\a\,,~~~~~~~~~~~\hat\chi_{\hat\b}={1\over 3}(\hat\l\g^p\hat\t)(\g_p\hat\t)_{\hat\b}}
We also introduce the Lorentz currents
\eqn\lorentz{L^{mn}=
 {1\over 2}: \p x^{[m} x^{n]} : +{1\over 2}:(p\g^{mn}\t): + : N^{mn}:\,,} 
 $$
\hat L^{pq}={1\over 2}:\bar\p x^{[p} x^{q]}:+{1\over 2}: (\hat p\g^{pq}\hat\t): + :\hat N^{pq}:\,,
$$
which satisfy the following commutation relations with the BRST charges
\eqn\lorentzQ{[Q_L,L^{mn}]=\p {\cal G}^{mn};~~~~~[Q_R,\hat L^{pq}]=\bar\p \hat {\cal G}^{pq}}
where 
\eqn\lorentzQII{{\cal G}^{mn}={1\over 4}(\t \g^r \l)\left(\d_r^{[m} x^{n]}+{1\over 4}(\t\g_r\g^{mn}\t)\right);~~~~~~~~~\hat {\cal G}^{pq}={1\over 4}(\hat\t \g^r \hat\l)\left(\d_r^{[p} x^{q]}+{1\over 4}(\hat\t\g_r\g^{pq}\hat\t\right)}
are BRST invariant.
By using the equations of motion from \conE\ it is easy to show that 
the currents $q_{\a}$, $\hat q_{\hat \b}$, $\l^{\a} d_{\a}$, $\hat \l^{\hat\b} \hat d_{\b}$, $L^{mn}$ and
$\hat L^{pq}$ are holomophic and anti-holomorphic, respectively. 

The energy-momentum tensor is given 
by 
\eqn\emt{
T_{zz} =- {1\over 2} \Pi^{m} \Pi_{m} - d_{\a} \p \t^{\a} - w_{\a} \p \l^{\a}\,, ~~~~~
\hat T_{\bar z\bar z} =- {1\over 2} \hat\Pi^{m} \hat \Pi_{m} -
\hat d_{\hat \a} \bar \p \hat \t^{\hat \a} - \hat w_{\hat \a} \bar\p \hat \l^{\hat \a}\,. 
}
where the last term in both expressions is invariant under the gauge transformations 
(A.4) which allow us to rewrite $T$ and $\hat T$ in terms of independent 
components of pure spinors. 

Our conventions for superspace covariant derivatives and 
supersymmetry charges are
\eqn\susu{\eqalign{
&D_{\a} = \p_{\a} + {1\over 2}(\g^{m}\t)_{\a} \p_{m}\,, ~~~~~~
Q_{\a} = \p_{\a} - {1\over 2}(\g^{m}\t)_{\a} \p_{m}\,, \cr
&\hat D_{\hat\a} = \p_{\hat\a} + {1\over 2}(\g^{m}\hat\t)_{\hat \a} \p_{m}\,, ~~~~~~
\hat Q_{\hat\a} = \p_{\hat\a} - {1\over 2}(\g^{m}\t)_{\hat\a} \p_{m}\,,
}
}
which satisfy
\eqn\deriv{\eqalign{
\left\{D_\a,D_\b\right\}=\g^m_{\a\b}\pa_m\,, \quad\quad
\left\{\hat D_{\hat\a},\hat D_{\hat\b}\right\}=\g^m_{\hat\a\hat\b}\pa_m\,, 
\quad\quad
\left\{D_\a,\hat D_{\hat\b}\right\}=0}}
$$
\{D_{\a}, Q_{\b} \} =0\,, ~~~~~
\{\hat D_{\hat\a},\hat Q_{\hat \b} \} =0\,.
$$


\appendix{B}{Solution of the Iterative Equations}

We list here the solution up to second order in both $\t^{\a}$ and $\hat\t^{\hat \a}$ 
for the superfields $A_{\a\hat\b}, A_{\a p}, A_{m \hat \b}, E_{\a}^{~~\hat\b}$ and $E^{\a}_{~~\hat \b}$. 

$$\eqalign{
A_{\a\hat \b}& = -{1\over 4}(\g^m\t)_\a(\g^p\hat\t)_{\hat\b}(g+b+\eta\phi)_{mp}\cr
&+{1\over 6}(\g^m\t)_\a(\g_m\t)_\g(\g^p\hat\t)_{\hat\b}\psi^\g_{~~p}+{1\over 6}(\g^m\t)_\a(\g^p\hat\t)_{\hat\b}(\g_p\hat\t)_{\hat\g}\psi_m^{~~\hat\g}\cr
&+{1\over 9}(\g^m\t)_\a (\g_m\t)_\b (\g^p\hat\t)_{\hat\b}(\g_p\hat\t)_{\hat\g}f^{\b\hat\g}+\dots \cr
& \cr
A_{\a p}& = {1 \over 2}\t^\b\g^m_{\b\a}(g+b+\eta\phi)_{mp} - {1\over 2}(\g^m\t)_\a(\g_p\hat\t)_{\hat\g}\psi_m^{~~\hat\g}+{1\over 3}(\g^m\t)_\a(\g_m\t)_\b\psi^\b_{~~p}\cr
&+{1\over 3}(\g^m\t)_\a(\g_m\t)_\b(\g_p\hat\t)_{\hat\g}f^{\b\hat\g}-{1\over 16}(\g^m\t)_\a(\g_p\hat\t)_{\hat\g}(\g^{qr}\hat\t)^{\hat\g}\omega_{m,qr}\cr
&-{1\over 24}(\g^m\t)_\a(\g_m\t)_\g(\g_p\hat\t)_{\hat\b}(\g^{qr}\hat\t)^{\hat\b}c^\g_{~~qr}+\dots \cr
& }$$
$$\eqalign{
A_{m \hat \b}& = -{1\over 2} \hat\t^{\hat\g}\g^p_{\hat\g\hat\b}(g+b+\eta\phi)_{mp} +{1\over 2}(\g_m\t)_\g(\g^p\hat\t)_{\hat\b}\psi^\g_{~~p}+ {1\over 3}(\g^p\hat\t)_{\hat\b}(\g_p\hat\t)_{\hat\g}\psi_m^{~~\hat\g}\cr
&+{1\over 16}(\g_m\t)_\g(\g^{nr}\t)^\g(\g^p\hat\t)_{\hat\b}\omega_{nr,p}+{1\over 3}(\g_m\t)_\b(\g^p\hat\t)_{\hat\b}(\g_p\hat\t)_{\hat\g}f^{\b\hat\g}\cr
&-{1\over 24}(\g_m\t)_\g(\g^{nr}\t)^\g (\g^p\hat\t)_{\hat\b}(\g_p\hat\t)_{\hat\g}c_{nr}^{~~~\hat\g}+\dots \cr
& 
}$$
$$\eqalign{
E_{\a}^{~~\hat\b}& = {1\over 2}\t^\g \g^m_{\g\a}\psi_m^{~~\hat\b}+{1\over 8} (\g^m\t)_\a (\g^{pq}\hat\t)^{\hat\b}\omega_{m,pq}+ {1\over 3}(\g^m\t)_\a(\g_m\t)_\g f^{\g\hat\b}\cr
&-{1\over 12}(\g^m\t)_\a(\g_m\t)_\g(\g^{pq}\hat\t)^{\hat\b}c^\g_{~~pq}+{1\over 8}(\g^m\t)_\a(\g^{pq}\hat\t)^{\hat\b}(\g_p\hat\t)_{\hat\g}\pa_q\psi_m^{~~\hat\g}\cr
& +{1\over 12}(\g^m\t)_\a (\g_m\t)_\g (\g^{pq}\hat\t)^{\hat\b}(\g_p\hat\t)_{\hat\g} \pa_q f^{\g\hat\g}+\dots\cr
& }$$
$$\eqalign{
E^{\a}_{~~\hat\b}& = {1\over 2}\hat\t^{\hat\g}\g^p_{\hat\g\hat\b}\psi^\a_{~~p}+ {1\over 8}(\g^{mn}\t)^\a(\g^p\hat\t)_{\hat\b}\omega_{mn,p}+ {1\over 3}(\g^p\hat\t)_{\hat\b}(\g_p\hat\t)_{\hat\g}f^{\a\hat\g}\cr
&-{1\over 8}(\g^{mn}\t)^\a(\g_m\t)_\g(\g^p\hat\t)_{\hat\b}\pa_n\psi^\g_{~~p}-{1\over 12}(\g^{mn}\t)^\a(\g^p\hat\t)_{\hat\b}(\g_p\hat\t)_{\hat\g}c_{mn}^{~~~\hat\g}\cr
&+{1\over 12}(\g^{mn}\t)^\a(\g_m\t)_\g (\g^p\hat\t)_{\hat\b}(\g_p\hat\t)_{\hat\g} \pa_n f^{\g\hat\g}+\dots
}$$


\appendix{C}{Solution of the Iterative Equations for Non-Constant RR Field-Strength}

Here we give the rest of the superfields for non linear $x$-dependent RR 
fields strengths.

$$\eqalign{ 
A_{\a p}& ={1\over 3}(\g^m\t)_\a(\g_m\t)_\b(\g_p\hat\t)_{\hat\g}(f^{\b\hat\g}+{\cal C}_n^{~~\b\hat\g}x^n) \cr
&+{1\over 36}(\g^m\t)_\a (\g_m\t)_\g(\g_p\hat\t)_{\hat\b}(\g^{qr}\hat\t)^{\hat\b}(\g_q \hat\t)_{\hat\g}{\cal C}_r^{~~\g\hat\g}\cr
&+{1\over 60}(\g^m\t)_\a(\g_m\t)_\b(\g^{nr}\t)^\b(\g_n\t)_\g(\g_p\hat\t)_{\hat\b}{\cal C}_r^{~~\g\hat\b}\cr
& }$$
$$\eqalign{ 
A_{m \hat \b}& = {1\over 3}(\g_m\t)_\b(\g^p\hat\t)_{\hat\b}(\g_p\hat\t)_{\hat\g}(f^{\b\hat\g}+{\cal C}_n^{~~\b\hat\g}x^n)\cr
&+{1\over 36}(\g_m\t)_\a(\g^{nr}\t)^\a(\g_n\t)_\g(\g_p\hat\t)_{\hat\b}(\g^p\hat\t)_{\hat\g}{\cal C}_r^{~~\g\hat\g}\cr
&+{1\over 60}(\g_m\t)_\g(\g^p\hat\t)_{\hat\b}(\g_p\hat\t)_{\hat\g}(\g^{rs}\hat\t)^{\hat\g}(\g_r\hat\t)_{\hat\d}{\cal C}_s^{~~\g\hat\d} \cr
& }$$ 
$$\eqalign{ 
E_{\a}^{~~\hat\b}& = {1\over 3}(\g^m\t)_\a(\g_m\t)_\g (f^{\g\hat\b}+{\cal C}_n^{~~\g\hat\b}x^n)\cr
& +{1\over 12}(\g^m\t)_\a (\g_m\t)_\g (\g^{pq}\hat\t)^{\hat\b}(\g_p\hat\t)_{\hat\g}{\cal C}_q^{~~\g\hat\g}+{1\over 60}(\g^m\t)_\a(\g_m\t)_\b(\g^{nr}\t)^\b(\g_n\t)_\g {\cal C}_r^{~~\g\hat\b}\cr
& }$$
$$\eqalign{ 
E^{\a}_{~~\hat\b}& = {1\over 3}(\g^p\hat\t)_{\hat\b}(\g_p\hat\t)_{\hat\g}(f^{\a\hat\g}+{\cal C}_m^{~~\a\hat\g}x^m)\cr
&+{1\over 12}(\g^{mn}\t)^\a(\g_m\t)_\g (\g^p\hat\t)_{\hat\b}(\g_p\hat\t)_{\hat\g}{\cal C}_n^{~~\g\hat\g}+{1\over 60}(\g^p\hat\t)_{\hat\b}(\g_p\hat\t)_{\hat\g}(\g^{qr}\hat\t)^{\hat\g}(\g_q\hat\t)_{\hat\d}{\cal C}_r^{~~\a\hat\d}\cr
}$$
$$
\eqalign{ 
A_{mp} &= (\g_m\t)_\b(\g_p\hat\t)_{\hat\g}(f^{\b\hat\g}+{\cal C}_n^{~~\b\hat\g}x^n) \cr
&+{1\over 12}(\g_m\t)_\b(\g^{nr}\t)^\b(\g_n\t)_\g(\g_p\hat\t)_{\hat\b}{\cal C}_r^{~~\g\hat\b}+{1\over 12}(\g_m\t)_\g(\g_p\hat\t)_{\hat\b}(\g^{rs}\hat\t)^{\hat\b}(\g_r\hat\t)_{\hat\g}{\cal C}_s^{~~\g\hat\g}\cr
& }$$
$$\eqalign{ 
E_{m}^{~~\hat\b} &= (\g_m\t)_\g (f^{\g\hat\b}+{\cal C}_n^{~~\g\hat\b}x^n)\cr
&+{1\over 4}(\g_m\t)_\g(\g^{pq}\hat\t)^{\hat\b}(\g_p\hat\t)_{\hat\g}{\cal C}_q^{~~\g\hat\g}+{1\over 12}(\g_m\t)_\a(\g^{nr}\t)^\a(\g_n\t)_\g {\cal C}_r^{~~\g\hat\b}\cr
& }$$
$$\eqalign{ 
E^{\a}_{~~p} &= (\g_p\hat\t)_{\hat\g}(f^{\a\hat\g}+{\cal C}_m^{~~\a\hat\g}x^m)\cr
&+{1\over 4}(\g^{mn}\t)^\a(\g_m\t)_\g (\g_p\hat\t)_{\hat\b}{\cal C}_n^{~~\g\hat\b}+{1\over 12}(\g_p\hat\t)_{\hat\b}(\g^{qr}\hat\t)^{\hat\b}(\g_q\hat\t)_{\hat\g}{\cal C}_r^{~~\a\hat\g}
\cr
& \cr
P^{\a\hat \b} &= (f^{\a\hat\b}+{\cal C}_m^{~~\a\hat\b}x^m)\cr
&+{1\over 4}(\g^{mn}\t)^\a (\g_m\t)_\g {\cal C}_n^{~~\g\hat\b} +{1\over 4}(\g^{pq}\hat\t)^{\hat\b}(\g_p\hat\t)_{\hat\g} {\cal C}_q^{~~\a\hat\g}\,.
}$$


\listrefs
\bye